\documentclass[manuscript=article]{achemso}

\usepackage[font=footnotesize,labelformat=simple]{subcaption}
\usepackage[T1]{fontenc}
\usepackage{amsmath, nccmath}
\usepackage{siunitx}

\usepackage{CJK}
\usepackage[colorlinks,bookmarks=false,citecolor=blue,linkcolor=red,urlcolor=blue]{hyperref}
\usepackage{color} 
\mciteErrorOnUnknownfalse
\usepackage{tikz}
\usepackage{overpic}
\usepackage{graphicx}
\usepackage{float}
\usepackage{multirow}
\usepackage{romannum}
\usepackage{bm}
\usepackage{amsfonts}
\usepackage{braket}
\usepackage{caption}
\usepackage{dblfloatfix}
\usepackage{cleveref}
\usepackage[font={scriptsize}]{caption}
\usepackage[version=3]{mhchem} 

\author{Rebecca Peake}
\affiliation{School of Physical and Chemical Sciences, Queen Mary University of London, London E1 4NS, United Kingdom}
\email{r.m.peake@qmul.ac.uk}

\author{Zo\'e Truyens}
\affiliation{Laboratory for Chemistry of Novel Materials, University of Mons, Mons, Belgium}

\author{Jan Mol}
\affiliation{School of Physical and Chemical Sciences, Queen Mary University of London, London E1 4NS, United Kingdom}

\author{Christian B Nielsen}
\affiliation{School of Physical and Chemical Sciences, Queen Mary University of London, London E1 4NS, United Kingdom}

\author{David Beljonne}
\affiliation{Laboratory for Chemistry of Novel Materials, University of Mons, Mons, Belgium}

\author{David Cornil}
\affiliation{Laboratory for Chemistry of Novel Materials, University of Mons, Mons, Belgium}

\author{Owen Benton}
\affiliation{School of Physical and Chemical Sciences, Queen Mary University of London, London E1 4NS, United Kingdom}
\email{j.o.benton@qmul.ac.uk}

\title{Engineering Topological Bands in Strained Covalent Organic Frameworks}

\begin{document}

\begin{abstract}
\noindent The tunability of covalent organic frameworks (COFs) opens opportunities to engineer topological electronic phases, including topological insulators (TIs) and higher-order topological insulators (HOTIs)—materials that host in-gap states localized at their edges, hinges, or corners. Here we explore how chemically feasible perturbations can drive triazine-based COFs (CTFs) into topological regimes. Using a tight-binding model on the Honeycomb lattice inspired by the frontier electronic states of CTFs, we show that introducing an effective uniaxial strain - implemented as a modulation of electron hopping on a subset of bonds - can generate a series of distinct topological band structures. This effect can be realized in practice through chemical substitution of linkers along the strained bonds. First-principles calculations demonstrate that replacing biphenyl with pyrene linkers drives a CTF to the brink of a HOTI phase, suggesting a viable route toward topological band-structure engineering in COFs.

\end{abstract}

\maketitle

\begin{figure}[H]
    \begin{subfigure}[b]{0.25\textwidth}
        \centering
        \includegraphics[width=\textwidth, trim={6cm 9cm 6cm 9cm}, clip]{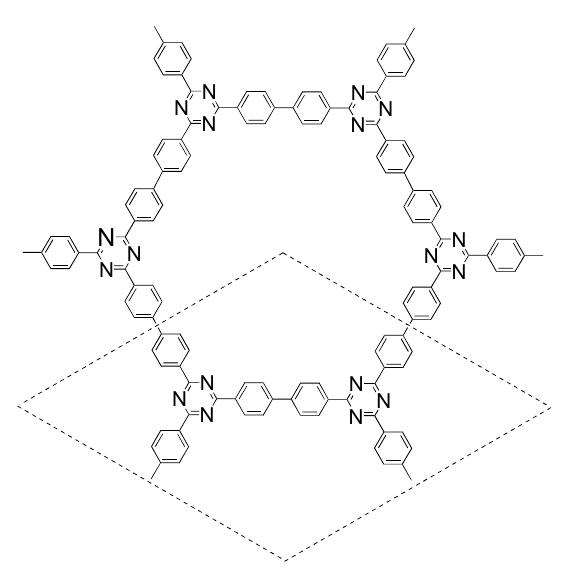}
        \caption{}
    \end{subfigure}%
    ~ 
    \begin{subfigure}[b]{0.25\textwidth}
            \begin{picture}(0,0)
        \put(60,20){\selectfont$D_{3}$}
    \end{picture}
        \includegraphics[width=0.42\textwidth, trim={10cm 13.2cm 10cm 13.2cm}, clip]{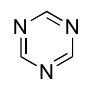}
        \centering
        \includegraphics[width=0.42\textwidth]{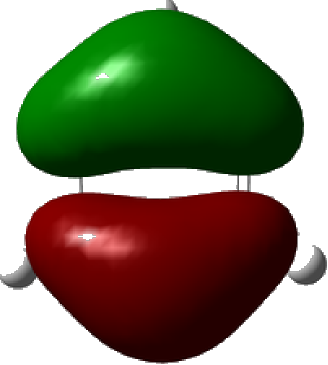}%
        ~
        \includegraphics[width=0.45\textwidth]{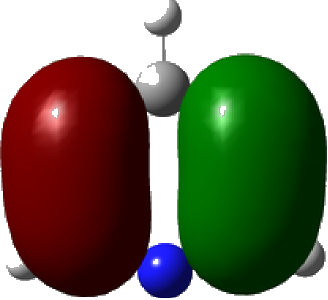}
        \caption{}
    \end{subfigure}
    
    \vskip\baselineskip

    \begin{subfigure}[b]{0.3\textwidth}
        \centering
        \includegraphics[width=1\textwidth,trim={8.6cm 12.8cm 8.6cm 12.8cm}, clip]{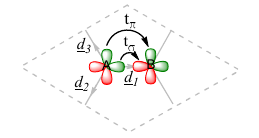}
        \caption{}
    \end{subfigure}%
    \begin{subfigure}[b]{0.4\textwidth}
        \centering
        \begin{overpic}[width=\textwidth]{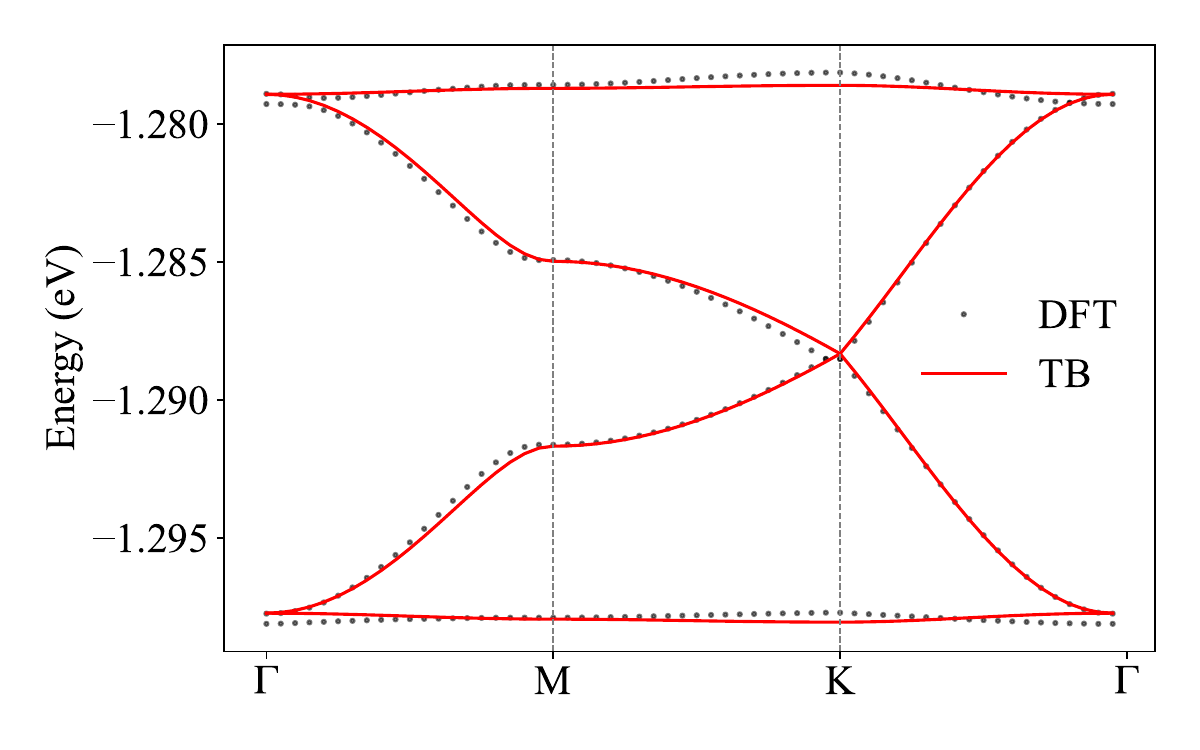}
          \put(19,19){\includegraphics[width=0.24\textwidth]{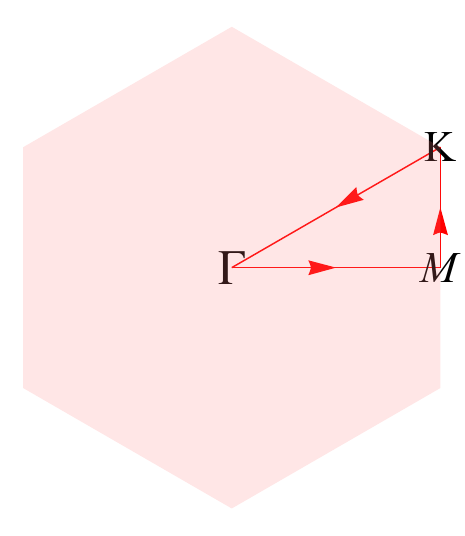}}
        \end{overpic}       
        \caption{}
    \end{subfigure}
    \caption{Tight-binding model description of a Honeycomb COF CTF-2 in the $\{p_{x},p_{y}\}$ basis. The dashed rhombus in (a) indicates the unit cell of CTF-2, with lattice parameters $a=b=22.06\text{\AA}$ and space group $P622$. (b) Shows the degenerate highest occupied MOs (HOMOs) of triazine. (c) Shows the unit cell of a Honeycomb lattice and illustrates hopping between degenerate orbitals of $\{p_{x},p_{y}\}$ character. (d) Displays the H-XY model (red) fitted to first-principles simulations of CTF-2, with parameters $t_{\sigma}=-0.108 {\rm meV}$, $t_{\pi}=0.637 {\rm meV}$, $E_0=-1.288 {\rm eV}$ about HOMO-3 - HOMO-6.}
    \label{fig:structure_and_orbitals}
\end{figure}
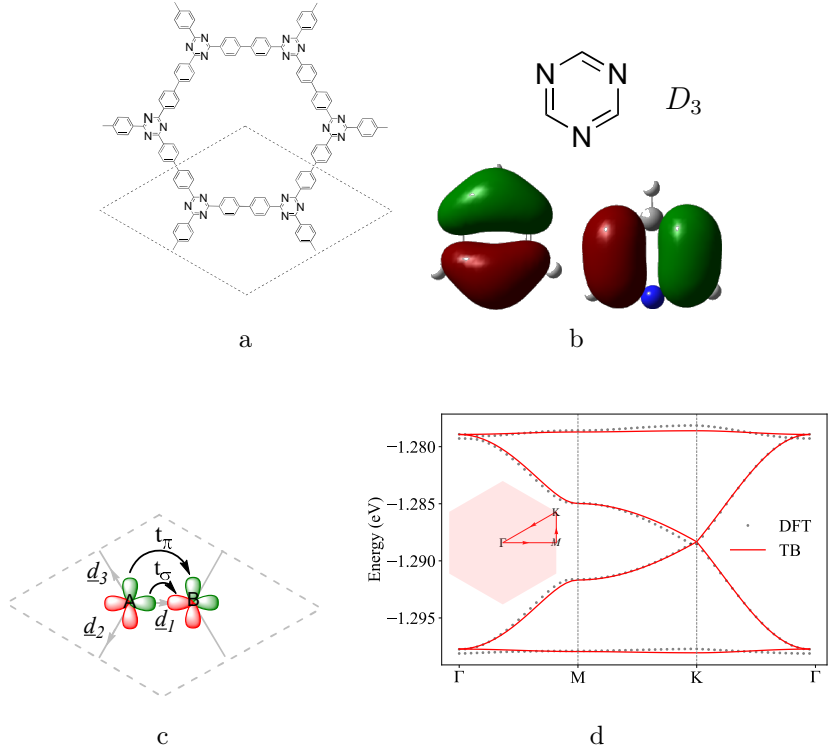

The search for materials that host topological electronic phases has become a central theme in condensed matter and materials chemistry \cite{Klitzing1980NewResistance,Thouless1955QuantizedPotential,Haldane1988ModelAnomaly,Bernevig2006QuantumWells,Koenig2007QuantumWells,Zhang2009TopologicalSurface,Hasan2010ColloquiumInsulators}, promising robust edge or corner states with potential applications in quantum technologies \cite{Mellnik2014Spin-transferInsulator}. In particular, higher-order topological insulators (HOTIs) have attracted intense interest as they extend the concept of bulk–boundary correspondence to lower-dimensional boundary modes \cite{Schindler2018Higher-orderInsulators, Benalcazar2017QuantizedInsulators,Schindler2018Higher-orderBismuth, Lee2023SpinfulWTe2}. Covalent organic frameworks (COFs), with their structural tunability and chemically designable building blocks, provide an ideal platform for realizing such topological phases through targeted molecular engineering \cite{Ni2024BandDesign, Springer2020TopologicalPolymers, Ni2022EngineeringCharacteristics}.
\noindent
Several 2D and 3D organic frameworks have been predicted to host HOTI states with 
corner and hinge modes \cite{Chen2024IntrinsicApproximation,Hu2022IntrinsicFrameworks, Ni2022OrganicFrameworks, Xue2021Higher-orderMaterials, Hu2023IdentifyingFrameworks}. Experimental verification is still awaited in most cases with metal-organic framework \ce{Ni3(HITP)2} being a notable confirmed example \cite{Hu2023IdentifyingFrameworks}. 
Two dimensional examples of HOTIs are of particular interest as quantum dots constructed from such systems can be employed for {\it cornertronics} - the manipulation of corner degrees of freedom using electric and optical fields \cite{Han2024CornertronicsInsulators}.
%
\noindent
Most theoretical discovery efforts for topological materials proceed by analysing {\it ab initio} band structures on a case-by-case basis \cite{Wang2013OrganicLattices,Wang2013PredictionInsulator,Liu2013FlatFramework,Wang2013QuantumInsulators,Ni2022EmergenceFramework,Ni2022OrganicFrameworks,Chen2024IntrinsicApproximation, Hu2023IdentifyingFrameworks, Hu2022IntrinsicFrameworks, Xue2021Higher-orderMaterials,Jiang2020TopologicalFrameworks,Gao2020QuantumLattice}. In this paper, we take a different approach, to explore the landscape of possibilities achievable by applying simple, chemically achievable, perturbations to existing COFs. 
\noindent
In particular, we consider a tight-binding model appropriate to the frontier states of a Covalent Triazine Framework (CTF), in which triazine nodes are connected by organic linkers \cite{M.LiuL.Guo2019CovalentApplications, K.WangL.Yang2017CovalentApproach, Kuhn2008PorousSynthesis}.
Fig. \ref{fig:structure_and_orbitals}(a) illustrates the structure of a particular CTF (CTF-2) where triazine cores are connected by biphenyl linkers \cite{Ren2012PorousSynthesis,Sun2022RapidFrameworks,Liu2019CovalentApplications, Chen2019DirectReactions, Kuhn2008MicroporousSynthesis, Jiang2022SynthesisBattery, Liao2023AdvancesFrameworks}.
The HOMOs on the triazine cores are two-fold degenerate and have the symmetry of $p_x$ and $p_y$ orbitals. 
We describe hopping between neighbouring sites
using the H-XY model \cite{Zhou2014EpitaxialSurface, Zhou2014FormationCoupling}:
\begin{equation}
H_{\rm XY}  =\sum_{\langle\mathbf{r},\mathbf{r'}\rangle} \sum_{\alpha,\beta} t_{\mathbf{r},\mathbf{r'}}^{\alpha,\beta}C_{\mathbf{r},\alpha}^{\dagger}C_{\mathbf{r'},\beta} + \it{H.C} + \sum_{\bf r}  \sum_{\alpha} E_0 C^{\dagger}_{{\bf r} \alpha} C_{{\bf r} \alpha}
\label{eq:HXY}
\end{equation}

\noindent
Here $C_{\mathbf{r},\alpha}^{(\dagger)}$ is the annihilation (creation) operator of an orbital, $\alpha$, at site $\mathbf{r}$.
\noindent
Due to the projection of the $\{p_{x},p_{y}\}$-like basis orbitals along different bonds, the hopping $t^{\alpha, \beta}_{{\bf r}, {\bf r}'}$ is bond-dependent.
$E_0$ is a constant, on-site, contribution to the energy.
\noindent
The orbital overlap between $p_x$ and $p_y$ orbitals is found using the Slater Koster integrals \cite{Slater1954SimplifiedProblem}.
This reduces the number of hopping parameters in the model to two: $t_{\sigma}$ and $t_{\pi}$ which describe the hopping between adjacent orbitals with relative orientation as in $\sigma$ or $\pi$ bonds. This is illustrated in Fig. \ref{fig:structure_and_orbitals}(c).
The third parameter, $E_0$ simply shifts
the bands by a constant energy.
After transforming to momentum space we arrive at a four band Hamiltonian, which produces the band structure shown in Fig. \ref{fig:structure_and_orbitals}(d).
\noindent
Fig. \ref{fig:structure_and_orbitals}(d) 
shows a comparison between a fit to the tight binding model Eq. (\ref{eq:HXY}) and first principles simulations of the band structure of CTF-2. 
Here we have selected a particular subset
of four bands which show substantial 
partial density of states (PDOS) on the triazine cores, and which are well described by the H-XY model. These bands
turn out to be a little below the Fermi level, but could be brought to the Fermi Level with a modest gate voltage or chemical doping.
The quality of the fit confirms that the H-XY model gives a good description of these bands (see Fig. \ref{fig:structure_and_orbitals}(d)).

\noindent
In addition to the two Dirac bands, from underlying Honeycomb symmetry as seen in Graphene, there exist two enclosing bands resulting from the extra orbital degrees of freedom \cite{Novoselov2004ElectricFilms}. If $t_{\pi}$ is small (as expected), these bands are nearly flat.
\noindent
In the absence of any further symmetry breaking or spin-orbit coupling (SOC), the four bands remain fully connected and the system is always 
 semi-metallic at half-filling \cite{Cano2018TopologyRepresentations, Bradlyn2019DisconnectedLattice,Ni2020-OrbitalFrameworks, Zhou2014FormationCoupling}.

\begin{figure}[H]
    \centering
    \begin{subfigure}[b]{0.22\textwidth}
        \centering
        \includegraphics[width=\textwidth,trim={8cm 11cm 8cm 11cm},clip]{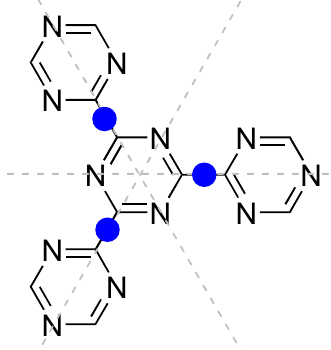}
        \caption{}
    \end{subfigure}%
    ~ 
    \begin{subfigure}[b]{0.22\textwidth}
        \centering
        \includegraphics[width=\textwidth,trim={8cm 11cm 8cm 11cm},clip]{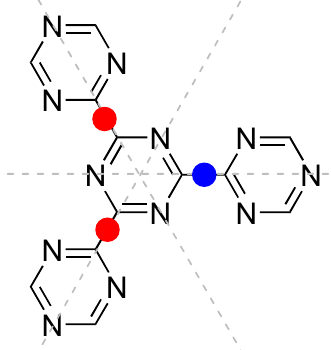}
        \caption{}
    \end{subfigure}%
    ~ 
    \begin{subfigure}[b]{0.3\textwidth}
        \centering
        \includegraphics[width=\textwidth,trim={6cm 11cm 6cm 11cm},clip]{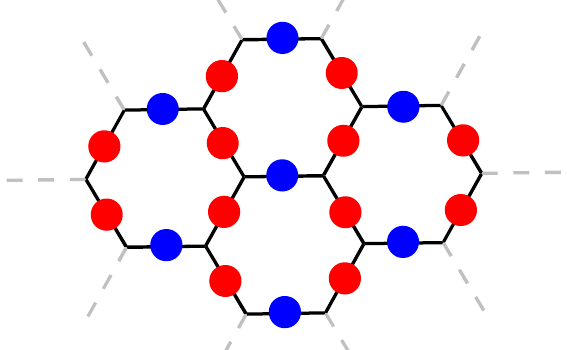}
        \caption{}
    \end{subfigure}%

    \caption{Structure of a CTF with (a) local $D_3$ and (b) local $C_2$ symmetry at the node (triazine) sites. Both are made up of triazine cores, where the blue(red) shapes represent different functional group linkers, with associated hopping strength $t_{\gamma}$($t_{\gamma_{0}}$) $\gamma \in \{\sigma, \pi\}$. (c) displays a larger fragment of the Honeycomb lattice made up from (b).}
    \label{fig:strained-lattice}
\end{figure}

\noindent
In order to access topologically insulating band structures, we need to open a gap. Since we are considering materials composed of light elements, we do not consider SOC as a mechanism for this. Instead, we consider simple ways to lower the lattice symmetry.
\noindent
A common choice of lattice distortion for this purpose is the Kekul{\'e} distortion on the Honeycomb lattice \cite{Chen2024IntrinsicApproximation,Qian2021Second-orderCandidates,Mu2022KekuleInsulator,Xue2021Higher-orderMaterials}. Here, we consider a simpler route - namely uniaxial strain. By straining the lattice as indicated in Fig. \ref{fig:strained-lattice} we can alter the space group from $P622$ to $C222$.
\noindent
Aside from mechanical strain, this change can be imposed by designing a Honeycomb framework material such that one of the three linkers connecting the cores is chemically different, as shown in Fig. \ref{fig:strained-lattice}. This {\it chemical} strain is uniquely achievable in organic framework materials, given their bottom-up design.

After this distortion, the tight-binding Hamiltonian maintains the form of Eq. (\ref{eq:HXY}), but now a third of the bonds have different values of $t_{\sigma}$ and $t_{\pi}$, as indicated in Fig. \ref{fig:strained-lattice}(c). We define the change in hopping parameter on the altered bonds such that  $t_{\gamma_{0}} = t_{\gamma}-\delta t_{\gamma}$ $\gamma \in \{\sigma,\pi\}$.
\noindent
Additionally, the change in on-site point group from $D_{3}$ to $C_{2}$ \cite{Mulliken1933ElectronicBond} splits the degeneracy between the $p_x$ and $p_y$ orbitals such that $E_{x}\ne E_{y}$, so we add an additional
orbital dependent chemical potential to the Hamiltonian
\begin{equation}
    H_{\rm XY} \to H_{\rm XY} + \sum_{\bf r} \delta E ( C_{{\bf r} x}^{\dagger} C_{{\bf r} x} - C_{{\bf r} y}^{\dagger} C_{{\bf r} y} )
\end{equation}
\noindent
We then explore the resulting band structures with varying $t_{\sigma_{0}}, t_{\pi_{0}}, \delta t_{\sigma}, \delta t_{\pi}$ and $\delta E$. The resulting phase diagram splits into different regions with gapped bands, separated by phase boundaries where bands touch. We must then decide which regions host topological band structures.
\noindent
For this classification task, we employ the methodology of Topological Quantum Chemistry (TQC) \cite{Bradlyn2017TopologicalChemistry, Cano2018BuildingRepresentations, Cano2020BandChemistry, Cano2018TopologyRepresentations,Kruthoff2017TopologicalCombinatorics,Bercioux2017Topological2017}. 
\noindent
The essence of this approach is to classify bands or collections of connected bands by the irreps of the little group of each high symmetry point of the Brillouin zone, which appear in the bands in question.
\noindent
If the collection of irreps thus obtained can be represented as a direct positive
sum of the elementary band representations obtainable from localized orbitals at Wyckoff positions in the unit cell, then the band structure represents an atomic insulator (i.e. it is trivial) because it can be continuously deformed to an atomic limit.
\noindent
Otherwise it must be topological. In this work we find \emph{fragile} topologically insulating bands, which remain topological unless combined with a trivial band \cite{Cano2020BandChemistry,Po2018FragileObstructions}. We consider the phase space of isolated fragile bands, which can be diagnosed via in-gap edge states.
\noindent
The case of an obstructed atomic insulator (OAI) occurs when the band(s) can be reduced to localized orbitals, but these orbitals would reside at
a Wyckoff position other than the actual Wyckoff position of the orbitals. OAIs are often associated with higher-order
topology and the existence of localized corner states -- and indeed we verify that the OAIs found in this work are HOTIs \cite{Benalcazar2017QuantizedInsulators}.
\noindent
Further details of our band structure classification are given in the 
Supporting Information.

\begin{figure}[H]
    \centering
    \begin{subfigure}[b]{0.3\textwidth}
        \centering
        \includegraphics[width=\textwidth]{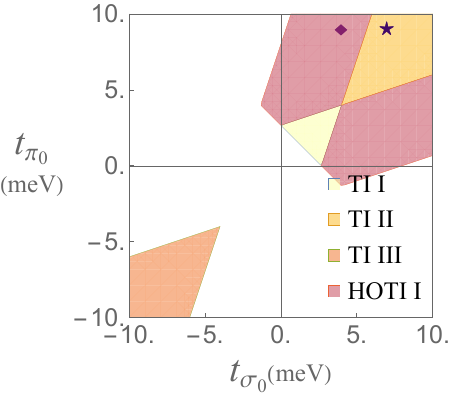}
        \caption{}
    \end{subfigure}%
    \hfill
    \centering
    \begin{subfigure}[b]{0.3\textwidth}
    \begin{overpic}[width=\textwidth]{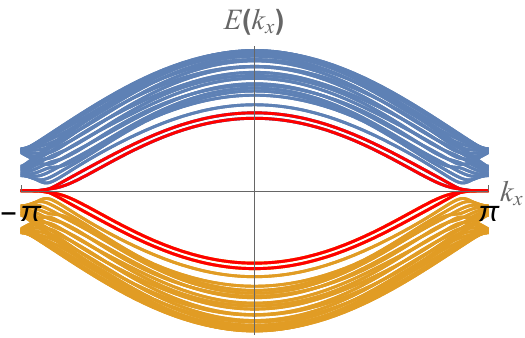}
        \put(80,40){\includegraphics[angle=90,width=0.2\textwidth]{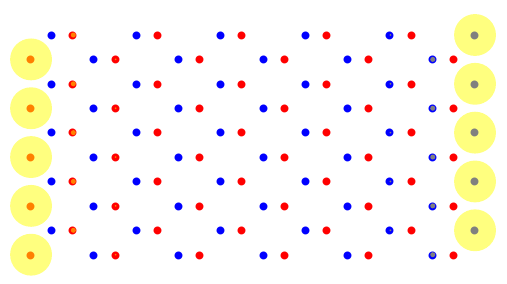}}
    \end{overpic}
    \caption{}
    \end{subfigure}%
    \hfill
    \begin{subfigure}[b]{0.3\textwidth}
    \centering
    \begin{overpic}[width=\textwidth]{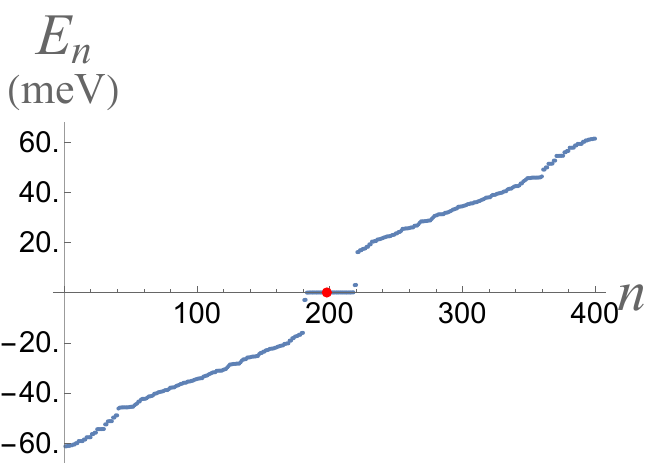}
        \put(20,40){\includegraphics[width=0.6\textwidth]{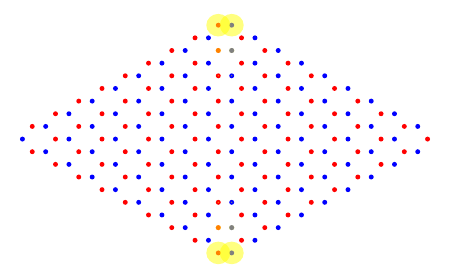}}
    \end{overpic}
    \caption{}
    \end{subfigure}%
    
    \caption{Variety of topological band structures achievable in the strained H-XY model. (a) Phase diagram showing varying $t_{\sigma_0}$ and $t_{\pi_0}$, with fixed  $\delta E=0\rm meV$ and $\delta t_{\pi}=\delta t_{\sigma}=-4\rm meV$. The coloured and white regions define topologically non-trivial (topologically insulating ($TI$) and higher-order topologically insulating ($HOTI$)) and trivial regions respectively. (b) Spectrum calculated on a nanoribbon with armchair edges for a parameter set in a TI phase (indicated by star on phase diagram), verifying the presence of gapless edge states at $\pm\pi$ as shown in the inset. (c) Spectrum calculated on a quantum dot with armchair edges for parameters in the HOTI regime (indicated by diamond on phase diagram). The in-gap states in the model of the spectrum are localized as the corners of the dot, as shown in inset which shows the square of wavefunction for one particular in-gap state (indicated by the red dot).}
\label{fig:phase_diagram_and_topology}    
\end{figure}

A two-dimensional slice of the phase diagram is shown in Fig. \ref{fig:phase_diagram_and_topology}(a). The white region in Fig. 
\ref{fig:phase_diagram_and_topology}(a) represents a trivial atomic insulator, while the coloured regions indicate TIs or HOTIs. 
The coloured region of parameter space shown in Fig.\ref{fig:phase_diagram_and_topology}(a) exhibits 3 TIs and 1 HOTI. By varying parameters, in total we found 4 distinct TIs and 2 distinct HOTIs defined in Eq. \ref{eq:TI} and \ref{eq:HOTI} respectively (see supporting information).
We have verified the topological character of each region of the phase 
diagram by calculating the spectrum on a nanoribbon armchair geometry
to identify edge states (in the cases of TIs) or on a quantum dot armchair 
geometry to identify corner states (in the case of HOTIs).

This establishes that a variety of topological band structures are accessible in principle by applying 
uniaxial strain to COFs like CTF-2.
We now turn to explore a specific 
implementation  of this.
We consider replacing the biphenyl linkers
on the bonds indicated in blue in Fig. \ref{fig:strained-lattice}
with pyrene.
We perform first-principles calculations
for such a COF by first relaxing the 
physical structure and then calculating the band structure \cite{M.J.FrischandG.W.TrucksandH.B.SchlegelandG.E.Scuseria2016GaussianC.01, RoyDenningtonandToddA.KeithandJohnM.Millam2019GaussView6,Kresse1996EfficientSet,Perdew1996GeneralizedSimple, Wang2021VASPKIT:Code}.
On the relaxation of the physical structure we find that it has space group
$C222$. 
Note that there are a number of space groups where action on $p_x$, $p_y$ orbitals is the same as (e.g. $Cmm2$) and our analysis would apply equally well to these COFs with these other space groups. These are enumerated in Tab. \ref{tab:spacegroups}.

\begin{figure}[H]
    \centering
    \begin{subfigure}[b]{0.5\textwidth}
        \centering
        \includegraphics[width=\textwidth]{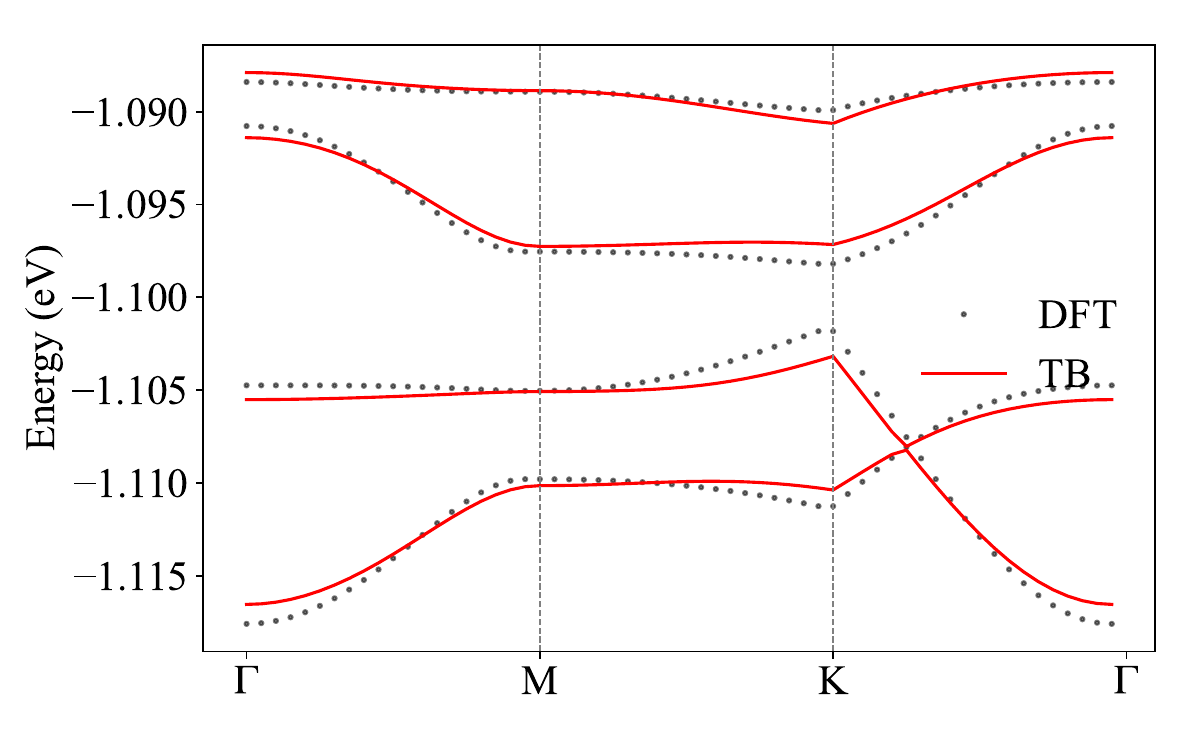}
        \caption{}
    \end{subfigure}%
    ~ 
    \begin{subfigure}[b]{0.4\textwidth}
        \centering
        \includegraphics[width=\textwidth]{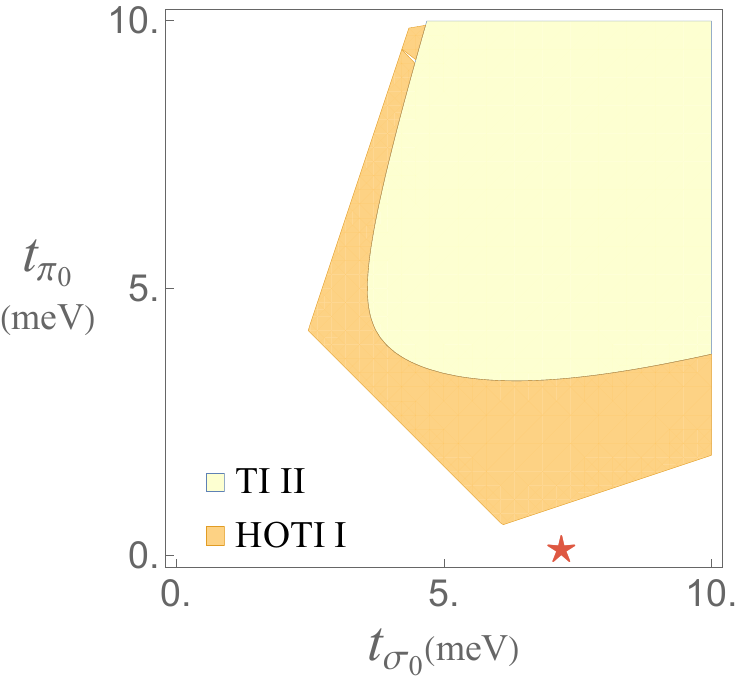}
        \caption{}
    \end{subfigure}
    \caption{Band structure and topological phase diagram for CTF-2 with replacement of biphenyl with pyrene linkers in one direction.
    (a) The predicted band structure, with a fit to the strained H-XY model with parameters $t_{\sigma_0}=7.20\rm meV$ and $t_{\pi_0}=0.121\rm meV$,
    (b) topological phase diagram with variation of $t_{\sigma_0}$ and $t_{\pi_0}$, with fixed values of $\delta E=-7.27\rm meV$, $\delta t_{\pi}=1.58\rm meV$ and $\delta t_{\sigma}=-2.18\rm meV$. The red star indicated the best fit parameters from simulations of strained CTF-2. This parameter set lies in the topologically trivial region, but on the border of a HOTI phase.}
    \label{fig:strained_bands_and_pd}
\end{figure}

%
As with the original framework, CTF-2, the bands relevant to the H-XY model turn out to be a little below the Fermi level, but can be brought to Fermi level with a  modest gate voltage.
A fit of the strained H-XY model to these bands is shown in Fig. \ref{fig:strained_bands_and_pd}(a), and agrees closely with the first principles
calculations.

Fig. \ref{fig:strained_bands_and_pd}(b)
shows a slice of the phase diagram of
the strained H-XY model, in the region
around the best fit parameters for strained CTF-2.
The best fit parameters fall into the 
topologically trivial (atomic insulator) regime. 
However, they lie very close to both
an HOTI and a TI. 

To be quantitative, about this closeness,
the overall bandwidth of the H-XY bands in the strained CTF-2 is predicted to be $\sim 23 \ {\rm meV}$. We can take this as representative of the overall scale of the hopping matrix elements. The variation in parameters needed to tip the system into the HOTI region is $\sim 1 \ {\rm meV}$. This variation corresponds to $4\%$ of the total bandwidth.

Thus, the particular chemically strained COF considered would require only a very small variation of the parameters to land in a HOTI phase, with localized fractional corner charges of size $e/2$.
A small further variation would bring the
system into a TI phase with gapless edge
modes.


In conclusion, we find that a variety of
topological states can be obtained from
existing Honeycomb COFs by using chemical substitutions that reproduce the effect of uniaxial strain.
This provides a simple, feasible and flexible mechanism to explore topological band structures, including HOTIs, in real systems.
We demonstrate the feasibility using ab-initio calculations for a particular COF with substitution of biphenyl linkers for pyrene, and find that the resulting structure is within a small parameter variation of a HOTI phase.
Given the variety of potential initial COFs and linker substitutions that could be made, this provides a simple route to likely realization of topological band structures in COFs.
Such COFs could in turn be used to create quantum dots, wherein localized corner states could be used as protected degrees of freedom for quantum sensing and manipulation of information.

More generally, our study underscores the usefulness of exploring the full phase diagram of realistic tight-binding models to identify routes to interesting phases and band structures.

\bibliography{CHP1,CHP2,CHP3,CHP4,SI}

\providecommand{\latin}[1]{#1}
\makeatletter
\providecommand{\doi}
  {\begingroup\let\do\@makeother\dospecials
  \catcode`\{=1 \catcode`\}=2 \doi@aux}
\providecommand{\doi@aux}[1]{\endgroup\texttt{#1}}
\makeatother
\providecommand*\mcitethebibliography{\thebibliography}
\csname @ifundefined\endcsname{endmcitethebibliography}
  {\let\endmcitethebibliography\endthebibliography}{}
\begin{mcitethebibliography}{62}
\providecommand*\natexlab[1]{#1}
\providecommand*\mciteSetBstSublistMode[1]{}
\providecommand*\mciteSetBstMaxWidthForm[2]{}
\providecommand*\mciteBstWouldAddEndPuncttrue
  {\def\EndOfBibitem{\unskip.}}
\providecommand*\mciteBstWouldAddEndPunctfalse
  {\let\EndOfBibitem\relax}
\providecommand*\mciteSetBstMidEndSepPunct[3]{}
\providecommand*\mciteSetBstSublistLabelBeginEnd[3]{}
\providecommand*\EndOfBibitem{}
\mciteSetBstSublistMode{f}
\mciteSetBstMaxWidthForm{subitem}{(\alph{mcitesubitemcount})}
\mciteSetBstSublistLabelBeginEnd
  {\mcitemaxwidthsubitemform\space}
  {\relax}
  {\relax}

\bibitem[Klitzing \latin{et~al.}(1980)Klitzing, Dorda, and
  Pepper]{Klitzing1980NewResistance}
Klitzing,~K.~V.; Dorda,~G.; Pepper,~M. {New Method for High-Accuracy
  Determination of the Fine-Structure Constant Based on Quantized Hall
  Resistance}. \emph{Phys. Rev. Lett.} \textbf{1980}, \emph{45}\relax
\mciteBstWouldAddEndPuncttrue
\mciteSetBstMidEndSepPunct{\mcitedefaultmidpunct}
{\mcitedefaultendpunct}{\mcitedefaultseppunct}\relax
\EndOfBibitem
\bibitem[Thouless \latin{et~al.}(1955)Thouless, Kohmoto, Nightingale, and den
  Nijs]{Thouless1955QuantizedPotential}
Thouless,~D.~J.; Kohmoto,~M.; Nightingale,~M.~P.; den Nijs,~M. {Quantized Hall
  Conductance in a Two-Dimensional Periodic Potential}. \emph{Ann. Phys.
  (N.Y.)} \textbf{1955}, \emph{49}, 1628\relax
\mciteBstWouldAddEndPuncttrue
\mciteSetBstMidEndSepPunct{\mcitedefaultmidpunct}
{\mcitedefaultendpunct}{\mcitedefaultseppunct}\relax
\EndOfBibitem
\bibitem[Haldane(1988)]{Haldane1988ModelAnomaly}
Haldane,~F.~D. {Model for a Quantum Hall Effect without Landau Levels:
  Condensed-Matter Realization of the "Parity Anomaly"}. \emph{Phys. Rev.
  Lett.} \textbf{1988}, \emph{61}, 2015\relax
\mciteBstWouldAddEndPuncttrue
\mciteSetBstMidEndSepPunct{\mcitedefaultmidpunct}
{\mcitedefaultendpunct}{\mcitedefaultseppunct}\relax
\EndOfBibitem
\bibitem[Bernevig \latin{et~al.}(2006)Bernevig, Hughes, and
  Zhang]{Bernevig2006QuantumWells}
Bernevig,~B.~A.; Hughes,~T.~L.; Zhang,~S.-C. {Quantum Spin Hall Effect and
  Topological Phase Transition in HgTe Quantum Wells}. \emph{Science}
  \textbf{2006}, \emph{314}, 1757--1761\relax
\mciteBstWouldAddEndPuncttrue
\mciteSetBstMidEndSepPunct{\mcitedefaultmidpunct}
{\mcitedefaultendpunct}{\mcitedefaultseppunct}\relax
\EndOfBibitem
\bibitem[Koenig \latin{et~al.}(2007)Koenig, Wiedmann, Bruene, Roth, Buhmann,
  Molenkamp, Qi, and Zhang]{Koenig2007QuantumWells}
Koenig,~M.; Wiedmann,~S.; Bruene,~C.; Roth,~A.; Buhmann,~H.; Molenkamp,~L.~W.;
  Qi,~X.-L.; Zhang,~S.-C. {Quantum Spin Hall Insulator State in HgTe Quantum
  Wells}. \emph{Science} \textbf{2007}, \emph{318}, 766--770\relax
\mciteBstWouldAddEndPuncttrue
\mciteSetBstMidEndSepPunct{\mcitedefaultmidpunct}
{\mcitedefaultendpunct}{\mcitedefaultseppunct}\relax
\EndOfBibitem
\bibitem[Zhang \latin{et~al.}(2009)Zhang, Liu, Qi, Dai, Fang, and
  Zhang]{Zhang2009TopologicalSurface}
Zhang,~H.; Liu,~C.~X.; Qi,~X.~L.; Dai,~X.; Fang,~Z.; Zhang,~S.~C. {Topological
  insulators in Bi 2 Se 3, Bi 2 Te 3 and Sb 2 Te 3 with a single Dirac cone on
  the surface}. \emph{Nat. Phys.} \textbf{2009}, \emph{5}, 438--442\relax
\mciteBstWouldAddEndPuncttrue
\mciteSetBstMidEndSepPunct{\mcitedefaultmidpunct}
{\mcitedefaultendpunct}{\mcitedefaultseppunct}\relax
\EndOfBibitem
\bibitem[Hasan and Kane(2010)Hasan, and Kane]{Hasan2010ColloquiumInsulators}
Hasan,~M.~Z.; Kane,~C.~L. {Colloquium Topological insulators}. \emph{Rev. Mod.
  Phys.} \textbf{2010}, \emph{82}, 3045\relax
\mciteBstWouldAddEndPuncttrue
\mciteSetBstMidEndSepPunct{\mcitedefaultmidpunct}
{\mcitedefaultendpunct}{\mcitedefaultseppunct}\relax
\EndOfBibitem
\bibitem[Mellnik \latin{et~al.}(2014)Mellnik, Lee, Richardella, Grab, Mintun,
  Fischer, Vaezi, Manchon, Kim, Samarth, and
  Ralph]{Mellnik2014Spin-transferInsulator}
Mellnik,~A.~R.; Lee,~J.~S.; Richardella,~A.; Grab,~J.~L.; Mintun,~P.~J.;
  Fischer,~M.~H.; Vaezi,~A.; Manchon,~A.; Kim,~E.-A.; Samarth,~N.; Ralph,~.
  D.~C. {Spin-transfer torque generated by a topological insulator}.
  \emph{Nature} \textbf{2014}, \emph{511}, 449--451\relax
\mciteBstWouldAddEndPuncttrue
\mciteSetBstMidEndSepPunct{\mcitedefaultmidpunct}
{\mcitedefaultendpunct}{\mcitedefaultseppunct}\relax
\EndOfBibitem
\bibitem[Schindler \latin{et~al.}(2018)Schindler, Cook, Vergniory, Wang,
  Parkin, Bernevig, and Neupert]{Schindler2018Higher-orderInsulators}
Schindler,~F.; Cook,~A.~M.; Vergniory,~M.~G.; Wang,~Z.; Parkin,~S. S.~P.;
  Bernevig,~B.~A.; Neupert,~T. {Higher-order topological insulators}.
  \emph{Sci. Adv.} \textbf{2018}, \emph{4}\relax
\mciteBstWouldAddEndPuncttrue
\mciteSetBstMidEndSepPunct{\mcitedefaultmidpunct}
{\mcitedefaultendpunct}{\mcitedefaultseppunct}\relax
\EndOfBibitem
\bibitem[Benalcazar \latin{et~al.}(2017)Benalcazar, Bernevig, and
  Hughes]{Benalcazar2017QuantizedInsulators}
Benalcazar,~W.~A.; Bernevig,~B.~A.; Hughes,~T.~L. {Quantized electric multipole
  insulators}. \emph{Science} \textbf{2017}, \emph{357}, 61--66\relax
\mciteBstWouldAddEndPuncttrue
\mciteSetBstMidEndSepPunct{\mcitedefaultmidpunct}
{\mcitedefaultendpunct}{\mcitedefaultseppunct}\relax
\EndOfBibitem
\bibitem[Schindler \latin{et~al.}(2018)Schindler, Wang, Vergniory, Cook,
  Murani, Sengupta, Kasumov, Deblock, Jeon, Drozdov, Bouchiat, Gu{\'{e}}ron,
  Yazdani, Bernevig, and Neupert]{Schindler2018Higher-orderBismuth}
Schindler,~F.; Wang,~Z.; Vergniory,~M.~G.; Cook,~A.~M.; Murani,~A.;
  Sengupta,~S.; Kasumov,~A.~Y.; Deblock,~R.; Jeon,~S.; Drozdov,~I.;
  Bouchiat,~H.; Gu{\'{e}}ron,~S.; Yazdani,~A.; Bernevig,~B.~A.; Neupert,~T.
  {Higher-order topology in bismuth}. \emph{Nat. Phys.} \textbf{2018},
  \emph{14}, 918--924\relax
\mciteBstWouldAddEndPuncttrue
\mciteSetBstMidEndSepPunct{\mcitedefaultmidpunct}
{\mcitedefaultendpunct}{\mcitedefaultseppunct}\relax
\EndOfBibitem
\bibitem[Lee \latin{et~al.}(2023)Lee, Kwon, Lee, Park, Cha, Watanabe,
  Taniguchi, Jo, and Choi]{Lee2023SpinfulWTe2}
Lee,~J.; Kwon,~J.; Lee,~E.; Park,~J.; Cha,~S.; Watanabe,~K.; Taniguchi,~T.;
  Jo,~M.~H.; Choi,~H. {Spinful hinge states in the higher-order topological
  insulators WTe2}. \emph{Nat. Commun.} \textbf{2023}, \emph{14}, 1--6\relax
\mciteBstWouldAddEndPuncttrue
\mciteSetBstMidEndSepPunct{\mcitedefaultmidpunct}
{\mcitedefaultendpunct}{\mcitedefaultseppunct}\relax
\EndOfBibitem
\bibitem[Ni and Br{\'{e}}das(2024)Ni, and Br{\'{e}}das]{Ni2024BandDesign}
Ni,~X.; Br{\'{e}}das,~J.-L. {Band engineering in two-dimensional porphyrin- and
  phthalocyanine-based covalent organic frameworks: insight from molecular
  design}. \emph{Moore and More} \textbf{2024}, \emph{1}\relax
\mciteBstWouldAddEndPuncttrue
\mciteSetBstMidEndSepPunct{\mcitedefaultmidpunct}
{\mcitedefaultendpunct}{\mcitedefaultseppunct}\relax
\EndOfBibitem
\bibitem[Springer \latin{et~al.}(2020)Springer, Liu, Kuc, and
  Heine]{Springer2020TopologicalPolymers}
Springer,~M.~A.; Liu,~T.~J.; Kuc,~A.; Heine,~T. {Topological two-dimensional
  polymers}. \emph{Chem. Soc. Rev.} \textbf{2020}, \emph{49}\relax
\mciteBstWouldAddEndPuncttrue
\mciteSetBstMidEndSepPunct{\mcitedefaultmidpunct}
{\mcitedefaultendpunct}{\mcitedefaultseppunct}\relax
\EndOfBibitem
\bibitem[Ni \latin{et~al.}(2022)Ni, Li, Liu, and
  Br{\'{e}}das]{Ni2022EngineeringCharacteristics}
Ni,~X.; Li,~H.; Liu,~F.; Br{\'{e}}das,~J.~L. {Engineering of flat bands and
  Dirac bands in two-dimensional covalent organic frameworks (COFs):
  Relationships among molecular orbital symmetry, lattice symmetry, and
  electronic-structure characteristics}. \emph{Mater. Horiz.} \textbf{2022},
  \emph{9}, 88--98\relax
\mciteBstWouldAddEndPuncttrue
\mciteSetBstMidEndSepPunct{\mcitedefaultmidpunct}
{\mcitedefaultendpunct}{\mcitedefaultseppunct}\relax
\EndOfBibitem
\bibitem[Chen \latin{et~al.}(2024)Chen, Xu, Xie, Xu, and
  Weng]{Chen2024IntrinsicApproximation}
Chen,~Z.; Xu,~S.; Xie,~Z.; Xu,~H.; Weng,~H. {Intrinsic second-order topological
  insulators in two-dimensional polymorphic graphyne with sublattice
  approximation}. \emph{npj Quantum Mater.} \textbf{2024}, \emph{9}\relax
\mciteBstWouldAddEndPuncttrue
\mciteSetBstMidEndSepPunct{\mcitedefaultmidpunct}
{\mcitedefaultendpunct}{\mcitedefaultseppunct}\relax
\EndOfBibitem
\bibitem[Hu \latin{et~al.}(2022)Hu, Zhang, Mu, and
  Wang]{Hu2022IntrinsicFrameworks}
Hu,~T.; Zhang,~T.; Mu,~H.; Wang,~Z. {Intrinsic Second-Order Topological
  Insulator in Two-Dimensional Covalent Organic Frameworks}. \emph{J. Phys.
  Chem. Lett.} \textbf{2022}, \emph{13}, 10905--10911\relax
\mciteBstWouldAddEndPuncttrue
\mciteSetBstMidEndSepPunct{\mcitedefaultmidpunct}
{\mcitedefaultendpunct}{\mcitedefaultseppunct}\relax
\EndOfBibitem
\bibitem[Ni \latin{et~al.}(2022)Ni, Huang, and
  Br{\'{e}}das]{Ni2022OrganicFrameworks}
Ni,~X.; Huang,~H.; Br{\'{e}}das,~J.~L. {Organic Higher-Order Topological
  Insulators: Heterotriangulene-Based Covalent Organic Frameworks}. \emph{J.
  Am. Chem. Soc.} \textbf{2022}, \emph{144}, 22778--22786\relax
\mciteBstWouldAddEndPuncttrue
\mciteSetBstMidEndSepPunct{\mcitedefaultmidpunct}
{\mcitedefaultendpunct}{\mcitedefaultseppunct}\relax
\EndOfBibitem
\bibitem[Xue \latin{et~al.}(2021)Xue, Huan, Zhao, Luo, Zhang, and
  Yang]{Xue2021Higher-orderMaterials}
Xue,~Y.; Huan,~H.; Zhao,~B.; Luo,~Y.; Zhang,~Z.; Yang,~Z. {Higher-order
  topological insulators in two-dimensional Dirac materials}. \emph{Phys. Rev.
  Research 3} \textbf{2021}, \emph{3}\relax
\mciteBstWouldAddEndPuncttrue
\mciteSetBstMidEndSepPunct{\mcitedefaultmidpunct}
{\mcitedefaultendpunct}{\mcitedefaultseppunct}\relax
\EndOfBibitem
\bibitem[Hu \latin{et~al.}(2023)Hu, Zhong, Zhang, Wang, and
  Wang]{Hu2023IdentifyingFrameworks}
Hu,~T.; Zhong,~W.; Zhang,~T.; Wang,~W.; Wang,~Z.~F. {Identifying topological
  corner states in two-dimensional metal-organic frameworks}. \emph{Nat.
  Commun.} \textbf{2023}, \emph{14}\relax
\mciteBstWouldAddEndPuncttrue
\mciteSetBstMidEndSepPunct{\mcitedefaultmidpunct}
{\mcitedefaultendpunct}{\mcitedefaultseppunct}\relax
\EndOfBibitem
\bibitem[Han \latin{et~al.}(2024)Han, Cui, Li, Zhang, Zhang, Yu, and
  Yao]{Han2024CornertronicsInsulators}
Han,~Y.; Cui,~C.; Li,~X.~P.; Zhang,~T.~T.; Zhang,~Z.; Yu,~Z.~M.; Yao,~Y.
  {Cornertronics in Two-Dimensional Second-Order Topological Insulators}.
  \emph{Phys. Rev. Lett.} \textbf{2024}, \emph{133}, 176602\relax
\mciteBstWouldAddEndPuncttrue
\mciteSetBstMidEndSepPunct{\mcitedefaultmidpunct}
{\mcitedefaultendpunct}{\mcitedefaultseppunct}\relax
\EndOfBibitem
\bibitem[Wang \latin{et~al.}(2013)Wang, Liu, and Liu]{Wang2013OrganicLattices}
Wang,~Z.~F.; Liu,~Z.; Liu,~F. {Organic topological insulators in organometallic
  lattices}. \emph{Nat. Commun.} \textbf{2013}, \emph{4}\relax
\mciteBstWouldAddEndPuncttrue
\mciteSetBstMidEndSepPunct{\mcitedefaultmidpunct}
{\mcitedefaultendpunct}{\mcitedefaultseppunct}\relax
\EndOfBibitem
\bibitem[Wang \latin{et~al.}(2013)Wang, Su, and
  Liu]{Wang2013PredictionInsulator}
Wang,~Z.~F.; Su,~N.; Liu,~F. {Prediction of a two-dimensional organic
  topological insulator}. \emph{Nano Lett.} \textbf{2013}, \emph{13},
  2842--2845\relax
\mciteBstWouldAddEndPuncttrue
\mciteSetBstMidEndSepPunct{\mcitedefaultmidpunct}
{\mcitedefaultendpunct}{\mcitedefaultseppunct}\relax
\EndOfBibitem
\bibitem[Liu \latin{et~al.}(2013)Liu, Wang, Mei, Wu, and
  Liu]{Liu2013FlatFramework}
Liu,~Z.; Wang,~Z.~F.; Mei,~J.~W.; Wu,~Y.~S.; Liu,~F. {Flat chern band in a
  two-dimensional organometallic framework}. \emph{Phys. Rev. Lett.}
  \textbf{2013}, \emph{110}\relax
\mciteBstWouldAddEndPuncttrue
\mciteSetBstMidEndSepPunct{\mcitedefaultmidpunct}
{\mcitedefaultendpunct}{\mcitedefaultseppunct}\relax
\EndOfBibitem
\bibitem[Wang \latin{et~al.}(2013)Wang, Liu, and
  Liu]{Wang2013QuantumInsulators}
Wang,~Z.~F.; Liu,~Z.; Liu,~F. {Quantum anomalous hall effect in 2D organic
  topological insulators}. \emph{Phys. Rev. Lett.} \textbf{2013},
  \emph{110}\relax
\mciteBstWouldAddEndPuncttrue
\mciteSetBstMidEndSepPunct{\mcitedefaultmidpunct}
{\mcitedefaultendpunct}{\mcitedefaultseppunct}\relax
\EndOfBibitem
\bibitem[Ni \latin{et~al.}(2022)Ni, Huang, and
  Br{\'{e}}das]{Ni2022EmergenceFramework}
Ni,~X.; Huang,~H.; Br{\'{e}}das,~J.~L. {Emergence of a Two-Dimensional
  Topological Dirac Semimetal Phase in a Phthalocyanine-Based Covalent Organic
  Framework}. \emph{Chem. Mater.} \textbf{2022}, \emph{34}, 3178--3184\relax
\mciteBstWouldAddEndPuncttrue
\mciteSetBstMidEndSepPunct{\mcitedefaultmidpunct}
{\mcitedefaultendpunct}{\mcitedefaultseppunct}\relax
\EndOfBibitem
\bibitem[Jiang \latin{et~al.}(2020)Jiang, Zhang, Wang, Liu, and
  Low]{Jiang2020TopologicalFrameworks}
Jiang,~W.; Zhang,~S.; Wang,~Z.; Liu,~F.; Low,~T. {Topological Band Engineering
  of Lieb Lattice in Phthalocyanine-Based Metal-Organic Frameworks}. \emph{Nano
  Lett.} \textbf{2020}, \emph{20}, 1959--1966\relax
\mciteBstWouldAddEndPuncttrue
\mciteSetBstMidEndSepPunct{\mcitedefaultmidpunct}
{\mcitedefaultendpunct}{\mcitedefaultseppunct}\relax
\EndOfBibitem
\bibitem[Gao \latin{et~al.}(2020)Gao, Zhang, Sun, Zhang, Zhang, and
  Du]{Gao2020QuantumLattice}
Gao,~Y.; Zhang,~Y.~Y.; Sun,~J.~T.; Zhang,~L.; Zhang,~S.; Du,~S. {Quantum
  anomalous Hall effect in two-dimensional Cu-dicyanobenzene coloring-triangle
  lattice}. \emph{Nano Research} \textbf{2020}, \emph{13}, 1571--1575\relax
\mciteBstWouldAddEndPuncttrue
\mciteSetBstMidEndSepPunct{\mcitedefaultmidpunct}
{\mcitedefaultendpunct}{\mcitedefaultseppunct}\relax
\EndOfBibitem
\bibitem[M.Liu~L.Guo(2019)]{M.LiuL.Guo2019CovalentApplications}
M.Liu~L.Guo,~S.~B. {Covalent triazine frameworks: Synthesis and applications}.
  \emph{J. Phys. Chem. A.} \textbf{2019}, \emph{7}\relax
\mciteBstWouldAddEndPuncttrue
\mciteSetBstMidEndSepPunct{\mcitedefaultmidpunct}
{\mcitedefaultendpunct}{\mcitedefaultseppunct}\relax
\EndOfBibitem
\bibitem[K.Wang~L.Yang(2017)]{K.WangL.Yang2017CovalentApproach}
K.Wang~L.Yang,~X. L. G. C. S. B.~A. {Covalent Triazine Frameworks via a
  Low-Temperature Polycondensation Approach}. \emph{Angew. Chem.}
  \textbf{2017}, \emph{129}, 14337--14341\relax
\mciteBstWouldAddEndPuncttrue
\mciteSetBstMidEndSepPunct{\mcitedefaultmidpunct}
{\mcitedefaultendpunct}{\mcitedefaultseppunct}\relax
\EndOfBibitem
\bibitem[Kuhn \latin{et~al.}(2008)Kuhn, Antonietti, and
  Thomas]{Kuhn2008PorousSynthesis}
Kuhn,~P.; Antonietti,~M.; Thomas,~A. {Porous, Covalent Triazine-Based
  Frameworks Prepared by Ionothermal Synthesis}. \emph{Angew. Chem. Int. Ed.}
  \textbf{2008}, \emph{47}, 3450--3453\relax
\mciteBstWouldAddEndPuncttrue
\mciteSetBstMidEndSepPunct{\mcitedefaultmidpunct}
{\mcitedefaultendpunct}{\mcitedefaultseppunct}\relax
\EndOfBibitem
\bibitem[Ren \latin{et~al.}(2012)Ren, Bojdys, Dawson, Laybourn, Khimyak, Adams,
  and Cooper]{Ren2012PorousSynthesis}
Ren,~S.; Bojdys,~M.~J.; Dawson,~R.; Laybourn,~A.; Khimyak,~Y.~Z.; Adams,~D.~J.;
  Cooper,~A.~I. {Porous, fluorescent, covalent triazine-based frameworks via
  room-temperature and microwave-assisted synthesis}. \emph{Advanced Materials}
  \textbf{2012}, \emph{24}, 2357--2361\relax
\mciteBstWouldAddEndPuncttrue
\mciteSetBstMidEndSepPunct{\mcitedefaultmidpunct}
{\mcitedefaultendpunct}{\mcitedefaultseppunct}\relax
\EndOfBibitem
\bibitem[Sun \latin{et~al.}(2022)Sun, Liang, and Xu]{Sun2022RapidFrameworks}
Sun,~T.; Liang,~Y.; Xu,~Y. {Rapid, Ordered Polymerization of Crystalline
  Semiconducting Covalent Triazine Frameworks}. \emph{Angew. Chem. Int. Ed.}
  \textbf{2022}, \emph{61}\relax
\mciteBstWouldAddEndPuncttrue
\mciteSetBstMidEndSepPunct{\mcitedefaultmidpunct}
{\mcitedefaultendpunct}{\mcitedefaultseppunct}\relax
\EndOfBibitem
\bibitem[Liu \latin{et~al.}(2019)Liu, Guo, Jin, and
  Tan]{Liu2019CovalentApplications}
Liu,~M.; Guo,~L.; Jin,~S.; Tan,~B. {Covalent triazine frameworks: Synthesis and
  applications}. \emph{J. Mat. Chem. A.} \textbf{2019}, \emph{7},
  5153--5172\relax
\mciteBstWouldAddEndPuncttrue
\mciteSetBstMidEndSepPunct{\mcitedefaultmidpunct}
{\mcitedefaultendpunct}{\mcitedefaultseppunct}\relax
\EndOfBibitem
\bibitem[Chen \latin{et~al.}(2019)Chen, Li, Hu, Hu, Liu, Yang, and
  Wen]{Chen2019DirectReactions}
Chen,~T.; Li,~W.-Q.; Hu,~W.-B.; Hu,~W.-J.; Liu,~Y.~A.; Yang,~H.; Wen,~K.
  {Direct synthesis of covalent triazine-based frameworks (CTFs) through
  aromatic nucleophilic substitution reactions}. \emph{RSC. Adv.}
  \textbf{2019}, \emph{9}\relax
\mciteBstWouldAddEndPuncttrue
\mciteSetBstMidEndSepPunct{\mcitedefaultmidpunct}
{\mcitedefaultendpunct}{\mcitedefaultseppunct}\relax
\EndOfBibitem
\bibitem[Kuhn \latin{et~al.}(2008)Kuhn, Antonietti, and
  Thomas]{Kuhn2008MicroporousSynthesis}
Kuhn,~P.; Antonietti,~M.; Thomas,~A. {Microporous Polymers Porous, Covalent
  Triazine-Based Frameworks Prepared by Ionothermal Synthesis}. \emph{Angew.
  Chem. Int. Ed.} \textbf{2008}, \emph{47}, 3450--3453\relax
\mciteBstWouldAddEndPuncttrue
\mciteSetBstMidEndSepPunct{\mcitedefaultmidpunct}
{\mcitedefaultendpunct}{\mcitedefaultseppunct}\relax
\EndOfBibitem
\bibitem[Jiang \latin{et~al.}(2022)Jiang, Wang, Qiu, Yang, Yang, Huang, Fang,
  and Li]{Jiang2022SynthesisBattery}
Jiang,~F.; Wang,~Y.; Qiu,~T.; Yang,~G.; Yang,~C.; Huang,~J.; Fang,~Z.; Li,~J.
  {Synthesis of biphenyl-linked covalent triazine frameworks with excellent
  lithium storage performance as anode in lithium ion battery}. \emph{J. Power
  Sources} \textbf{2022}, \emph{523}, 231041\relax
\mciteBstWouldAddEndPuncttrue
\mciteSetBstMidEndSepPunct{\mcitedefaultmidpunct}
{\mcitedefaultendpunct}{\mcitedefaultseppunct}\relax
\EndOfBibitem
\bibitem[Liao \latin{et~al.}(2023)Liao, Li, Yin, Chen, Zhong, Du, Liu, He, Fu,
  and Zeng]{Liao2023AdvancesFrameworks}
Liao,~L.; Li,~M.; Yin,~Y.; Chen,~J.; Zhong,~Q.; Du,~R.; Liu,~S.; He,~Y.;
  Fu,~W.; Zeng,~F. {Advances in the Synthesis of Covalent Triazine Frameworks}.
  \emph{ACS Omega} \textbf{2023}, \emph{8}, 4527--4542\relax
\mciteBstWouldAddEndPuncttrue
\mciteSetBstMidEndSepPunct{\mcitedefaultmidpunct}
{\mcitedefaultendpunct}{\mcitedefaultseppunct}\relax
\EndOfBibitem
\bibitem[Zhou \latin{et~al.}(2014)Zhou, Ming, Liu, Wang, Li, and
  Liu]{Zhou2014EpitaxialSurface}
Zhou,~M.; Ming,~W.; Liu,~Z.; Wang,~Z.; Li,~P.; Liu,~F. {Epitaxial growth of
  large-gap quantum spin Hall insulator on semiconductor surface}. \emph{Proc.
  Natl. Acad. Sci. U.S.A.} \textbf{2014}, \emph{111}, 14378--14381\relax
\mciteBstWouldAddEndPuncttrue
\mciteSetBstMidEndSepPunct{\mcitedefaultmidpunct}
{\mcitedefaultendpunct}{\mcitedefaultseppunct}\relax
\EndOfBibitem
\bibitem[Zhou \latin{et~al.}(2014)Zhou, Ming, Liu, Wang, Yao, and
  Liu]{Zhou2014FormationCoupling}
Zhou,~M.; Ming,~W.; Liu,~Z.; Wang,~Z.; Yao,~Y.; Liu,~F. {Formation of quantum
  spin Hall state on Si surface and energy gap scaling with strength of spin
  orbit coupling}. \emph{Sci. Rep.} \textbf{2014}, \emph{4}\relax
\mciteBstWouldAddEndPuncttrue
\mciteSetBstMidEndSepPunct{\mcitedefaultmidpunct}
{\mcitedefaultendpunct}{\mcitedefaultseppunct}\relax
\EndOfBibitem
\bibitem[Slater and Koster(1954)Slater, and
  Koster]{Slater1954SimplifiedProblem}
Slater,~J.~C.; Koster,~G.~F. {Simplified LCAO Method for the Periodic Potential
  Problem}. \emph{Phys. Rev.} \textbf{1954}, \emph{94}, 1498--1524\relax
\mciteBstWouldAddEndPuncttrue
\mciteSetBstMidEndSepPunct{\mcitedefaultmidpunct}
{\mcitedefaultendpunct}{\mcitedefaultseppunct}\relax
\EndOfBibitem
\bibitem[Novoselov \latin{et~al.}(2004)Novoselov, Geim, Morozov, Jiang, Zhang,
  Dubonos, Grigorieva, and Firsov]{Novoselov2004ElectricFilms}
Novoselov,~K.~S.; Geim,~A.~K.; Morozov,~S.~V.; Jiang,~D.; Zhang,~Y.;
  Dubonos,~S.~V.; Grigorieva,~I.~V.; Firsov,~A.~A. {Electric Field Effect in
  Atomically Thin Carbon Films}. \emph{Science} \textbf{2004}, \emph{306},
  666--669\relax
\mciteBstWouldAddEndPuncttrue
\mciteSetBstMidEndSepPunct{\mcitedefaultmidpunct}
{\mcitedefaultendpunct}{\mcitedefaultseppunct}\relax
\EndOfBibitem
\bibitem[Cano \latin{et~al.}(2018)Cano, Bradlyn, Wang, Elcoro, Vergniory,
  Felser, Aroyo, and Bernevig]{Cano2018TopologyRepresentations}
Cano,~J.; Bradlyn,~B.; Wang,~Z.; Elcoro,~L.; Vergniory,~M.~G.; Felser,~C.;
  Aroyo,~M.~I.; Bernevig,~B.~A. {Topology of Disconnected Elementary Band
  Representations}. \emph{Phys. Rev. Lett.} \textbf{2018}, \emph{120}\relax
\mciteBstWouldAddEndPuncttrue
\mciteSetBstMidEndSepPunct{\mcitedefaultmidpunct}
{\mcitedefaultendpunct}{\mcitedefaultseppunct}\relax
\EndOfBibitem
\bibitem[Bradlyn \latin{et~al.}(2019)Bradlyn, Wang, Cano, and
  Bernevig]{Bradlyn2019DisconnectedLattice}
Bradlyn,~B.; Wang,~Z.; Cano,~J.; Bernevig,~B.~A. {Disconnected elementary band
  representations, fragile topology, and Wilson loops as topological indices:
  An example on the triangular lattice}. \emph{Phys. Rev. B} \textbf{2019},
  \emph{99}\relax
\mciteBstWouldAddEndPuncttrue
\mciteSetBstMidEndSepPunct{\mcitedefaultmidpunct}
{\mcitedefaultendpunct}{\mcitedefaultseppunct}\relax
\EndOfBibitem
\bibitem[Ni \latin{et~al.}(2020)Ni, Zhou, Sethi, and
  Liu]{Ni2020-OrbitalFrameworks}
Ni,~X.; Zhou,~Y.; Sethi,~G.; Liu,~F. {{$\pi$}-Orbital Yin-Yang Kagome bands in
  anilato-based metal-organic frameworks}. \emph{Phys. Chem. Chem. Phys.}
  \textbf{2020}, \emph{22}, 25827--25832\relax
\mciteBstWouldAddEndPuncttrue
\mciteSetBstMidEndSepPunct{\mcitedefaultmidpunct}
{\mcitedefaultendpunct}{\mcitedefaultseppunct}\relax
\EndOfBibitem
\bibitem[Qian \latin{et~al.}(2021)Qian, Liu, and
  Yao]{Qian2021Second-orderCandidates}
Qian,~S.; Liu,~C.~C.; Yao,~Y. {Second-order topological insulator state in
  hexagonal lattices and its abundant material candidates}. \emph{Phys. Rev. B}
  \textbf{2021}, \emph{104}, 245427\relax
\mciteBstWouldAddEndPuncttrue
\mciteSetBstMidEndSepPunct{\mcitedefaultmidpunct}
{\mcitedefaultendpunct}{\mcitedefaultseppunct}\relax
\EndOfBibitem
\bibitem[Mu \latin{et~al.}(2022)Mu, Liu, Hu, and Wang]{Mu2022KekuleInsulator}
Mu,~H.; Liu,~B.; Hu,~T.; Wang,~Z. {Kekul{\'{e}} Lattice in Graphdiyne:
  Coexistence of Phononic and Electronic Second-Order Topological Insulator}.
  \emph{Nano Lett.} \textbf{2022}, \emph{22}, 1122--1128\relax
\mciteBstWouldAddEndPuncttrue
\mciteSetBstMidEndSepPunct{\mcitedefaultmidpunct}
{\mcitedefaultendpunct}{\mcitedefaultseppunct}\relax
\EndOfBibitem
\bibitem[Mulliken(1933)]{Mulliken1933ElectronicBond}
Mulliken,~R.~S. {Electronic Structures of Polyatomic Molecules and Valence. IV.
  Electronic States, Quantum Theory of the Double Bond}. \emph{Phys. Rev.}
  \textbf{1933}, \emph{43}, 279\relax
\mciteBstWouldAddEndPuncttrue
\mciteSetBstMidEndSepPunct{\mcitedefaultmidpunct}
{\mcitedefaultendpunct}{\mcitedefaultseppunct}\relax
\EndOfBibitem
\bibitem[Bradlyn \latin{et~al.}(2017)Bradlyn, Elcoro, Cano, Vergniory, Wang,
  Felser, Aroyo, and Bernevig]{Bradlyn2017TopologicalChemistry}
Bradlyn,~B.; Elcoro,~L.; Cano,~J.; Vergniory,~M.~G.; Wang,~Z.; Felser,~C.;
  Aroyo,~M.~I.; Bernevig,~B.~A. {Topological quantum chemistry}. \emph{Nature}
  \textbf{2017}, \emph{547}, 298--305\relax
\mciteBstWouldAddEndPuncttrue
\mciteSetBstMidEndSepPunct{\mcitedefaultmidpunct}
{\mcitedefaultendpunct}{\mcitedefaultseppunct}\relax
\EndOfBibitem
\bibitem[Cano \latin{et~al.}(2018)Cano, Bradlyn, Wang, Elcoro, Vergniory,
  Felser, Aroyo, and Bernevig]{Cano2018BuildingRepresentations}
Cano,~J.; Bradlyn,~B.; Wang,~Z.; Elcoro,~L.; Vergniory,~M.~G.; Felser,~C.;
  Aroyo,~M.~I.; Bernevig,~B.~A. {Building blocks of topological quantum
  chemistry: Elementary band representations}. \emph{Phys. Rev. B.}
  \textbf{2018}, \emph{97}\relax
\mciteBstWouldAddEndPuncttrue
\mciteSetBstMidEndSepPunct{\mcitedefaultmidpunct}
{\mcitedefaultendpunct}{\mcitedefaultseppunct}\relax
\EndOfBibitem
\bibitem[Cano and Bradlyn(2020)Cano, and Bradlyn]{Cano2020BandChemistry}
Cano,~J.; Bradlyn,~B. {Band Representations and Topological Quantum Chemistry}.
  \emph{Annu. Rev. Condens. Matter Phys.} \textbf{2020}, \emph{12},
  225--246\relax
\mciteBstWouldAddEndPuncttrue
\mciteSetBstMidEndSepPunct{\mcitedefaultmidpunct}
{\mcitedefaultendpunct}{\mcitedefaultseppunct}\relax
\EndOfBibitem
\bibitem[Kruthoff \latin{et~al.}(2017)Kruthoff, De~Boer, Van~Wezel, Kane, and
  Slager]{Kruthoff2017TopologicalCombinatorics}
Kruthoff,~J.; De~Boer,~J.; Van~Wezel,~J.; Kane,~C.~L.; Slager,~R.~J.
  {Topological classification of crystalline insulators through band structure
  combinatorics}. \emph{Phys. Rev. X} \textbf{2017}, \emph{7}\relax
\mciteBstWouldAddEndPuncttrue
\mciteSetBstMidEndSepPunct{\mcitedefaultmidpunct}
{\mcitedefaultendpunct}{\mcitedefaultseppunct}\relax
\EndOfBibitem
\bibitem[Bercioux \latin{et~al.}(2017)Bercioux, Cayssol, Vergniory, and
  Calvo]{Bercioux2017Topological2017}
Bercioux,~D.; Cayssol,~J.; Vergniory,~M.~G.; Calvo,~M.~R. {Topological Matter
  Lectures from the Topological Matter School 2017}. 2017\relax
\mciteBstWouldAddEndPuncttrue
\mciteSetBstMidEndSepPunct{\mcitedefaultmidpunct}
{\mcitedefaultendpunct}{\mcitedefaultseppunct}\relax
\EndOfBibitem
\bibitem[Po \latin{et~al.}(2018)Po, Watanabe, and
  Vishwanath]{Po2018FragileObstructions}
Po,~H.~C.; Watanabe,~H.; Vishwanath,~A. {Fragile Topology and Wannier
  Obstructions}. \emph{Phys. Rev. Lett} \textbf{2018}, \emph{121}, 126402\relax
\mciteBstWouldAddEndPuncttrue
\mciteSetBstMidEndSepPunct{\mcitedefaultmidpunct}
{\mcitedefaultendpunct}{\mcitedefaultseppunct}\relax
\EndOfBibitem
\bibitem[{M. J. Frisch and G. W. Trucks and H. B. Schlegel and G. E. Scuseria}
  \latin{et~al.}(2016){M. J. Frisch and G. W. Trucks and H. B. Schlegel and G.
  E. Scuseria}, {M. A. Robb and J. R. Cheeseman and G. Scalmani and V. Barone
  and G. A. Petersson and H. Nakatsuji}, {X. Li and M. Caricato and A. V.
  Marenich and J. Bloino and B. G. Janesko and R. Gomperts}, {B. Mennucci and
  H. P. Hratchian and J. V. Ortiz and A. F. Izmaylov and J. L. Sonnenberg}, {D.
  Williams-Young and F. Ding and F. Lipparini and F. Egidi and J. Goings and B.
  Peng}, {A. Petrone and T. Henderson and D. Ranasinghe and V. G. Zakrzewski
  and J. Gao and N. Rega}, {G. Zheng and W. Liang and M. Hada and M. Ehara and
  K. Toyota and R. Fukuda and J. Hasegawa and M. Ishida and T. Nakajima and Y.
  Honda and O. Kitao and H. Nakai and T. Vreven}, Throssell, Montgomery,
  Peralta, Ogliaro, Bearpark, Heyd, Brothers, Kudin, Staroverov, Keith, {R.
  Kobayashi and J. Normand and K. Raghavachari and A. P. Rendell and J. C.
  Burant and S. S. Iyengar and J. Tomasi and M. Cossi and J. M. Millam and M.
  Klene and C. Adamo}, and {R. Cammi and J. W. Ochterski and R. L. Martin and
  K. Morokuma and O. Farkas and J. B. Foresman and D. J.
  Fox}]{M.J.FrischandG.W.TrucksandH.B.SchlegelandG.E.Scuseria2016GaussianC.01}
{M. J. Frisch and G. W. Trucks and H. B. Schlegel and G. E. Scuseria}
  \latin{et~al.}  {Gaussian 16 Revision C.01}. 2016\relax
\mciteBstWouldAddEndPuncttrue
\mciteSetBstMidEndSepPunct{\mcitedefaultmidpunct}
{\mcitedefaultendpunct}{\mcitedefaultseppunct}\relax
\EndOfBibitem
\bibitem[{Roy Dennington and Todd A. Keith and John M.
  Millam}(2019)]{RoyDenningtonandToddA.KeithandJohnM.Millam2019GaussView6}
{Roy Dennington and Todd A. Keith and John M. Millam} {GaussView Version 6}.
  2019\relax
\mciteBstWouldAddEndPuncttrue
\mciteSetBstMidEndSepPunct{\mcitedefaultmidpunct}
{\mcitedefaultendpunct}{\mcitedefaultseppunct}\relax
\EndOfBibitem
\bibitem[Kresse and Furthm{\"{u}}ller(1996)Kresse, and
  Furthm{\"{u}}ller]{Kresse1996EfficientSet}
Kresse,~G.; Furthm{\"{u}}ller,~J. {Efficient iterative schemes for ab initio
  total-energy calculations using a plane-wave basis set}. \emph{Phys. Rev. B}
  \textbf{1996}, \emph{54}, 11169\relax
\mciteBstWouldAddEndPuncttrue
\mciteSetBstMidEndSepPunct{\mcitedefaultmidpunct}
{\mcitedefaultendpunct}{\mcitedefaultseppunct}\relax
\EndOfBibitem
\bibitem[Perdew \latin{et~al.}(1996)Perdew, Burke, and
  Ernzerhof]{Perdew1996GeneralizedSimple}
Perdew,~J.~P.; Burke,~K.; Ernzerhof,~M. {Generalized Gradient Approximation
  Made Simple}. \emph{Phys. Rev. Lett.} \textbf{1996}, \emph{77}, 3865\relax
\mciteBstWouldAddEndPuncttrue
\mciteSetBstMidEndSepPunct{\mcitedefaultmidpunct}
{\mcitedefaultendpunct}{\mcitedefaultseppunct}\relax
\EndOfBibitem
\bibitem[Wang \latin{et~al.}(2021)Wang, Xu, Liu, Tang, and
  Geng]{Wang2021VASPKIT:Code}
Wang,~V.; Xu,~N.; Liu,~J.~C.; Tang,~G.; Geng,~W.~T. {VASPKIT: A user-friendly
  interface facilitating high-throughput computing and analysis using VASP
  code}. \emph{Comput. Phys. Commun.} \textbf{2021}, \emph{267}, 108033\relax
\mciteBstWouldAddEndPuncttrue
\mciteSetBstMidEndSepPunct{\mcitedefaultmidpunct}
{\mcitedefaultendpunct}{\mcitedefaultseppunct}\relax
\EndOfBibitem
\bibitem[Aroyo \latin{et~al.}(2006)Aroyo, Perez-Mato, Capillas, Kroumova,
  Ivantchev, Madariaga, Kirov, and Wondratschek]{Aroyo2006BilbaoPrograms}
Aroyo,~M.~I.; Perez-Mato,~J.~M.; Capillas,~C.; Kroumova,~E.; Ivantchev,~S.;
  Madariaga,~G.; Kirov,~A.; Wondratschek,~H. {Bilbao Crystallographic Server:
  I. Databases and crystallographic computing programs}. \emph{Zeitschrift fur
  Kristallographie} \textbf{2006}, \emph{221}, 15--27\relax
\mciteBstWouldAddEndPuncttrue
\mciteSetBstMidEndSepPunct{\mcitedefaultmidpunct}
{\mcitedefaultendpunct}{\mcitedefaultseppunct}\relax
\EndOfBibitem
\bibitem[Aroyo \latin{et~al.}(2006)Aroyo, Kirov, Capillas, Perez-Mato, and
  Wondratschek]{Aroyo2006BilbaoGroups}
Aroyo,~M.~I.; Kirov,~A.; Capillas,~C.; Perez-Mato,~J.~M.; Wondratschek,~H.
  {Bilbao Crystallographic Server. II. Representations of crystallographic
  point groups and space groups}. \emph{Acta. Crystallogr. A Found. Adv.}
  \textbf{2006}, \emph{62}, 115--128\relax
\mciteBstWouldAddEndPuncttrue
\mciteSetBstMidEndSepPunct{\mcitedefaultmidpunct}
{\mcitedefaultendpunct}{\mcitedefaultseppunct}\relax
\EndOfBibitem
\end{mcitethebibliography}

\newpage
\section{Supplementary Information}
\setcounter{figure}{0}
\renewcommand{\thefigure}{S\arabic{figure}}

\setcounter{table}{0}
\renewcommand{\thetable}{S\arabic{table}}

\setcounter{equation}{0}
\renewcommand{\theequation}{S\arabic{equation}}

\section{Appendix A: Symmetry Analysis}

\begin{figure}[H]
    \centering
    \includegraphics[width=0.4\textwidth]{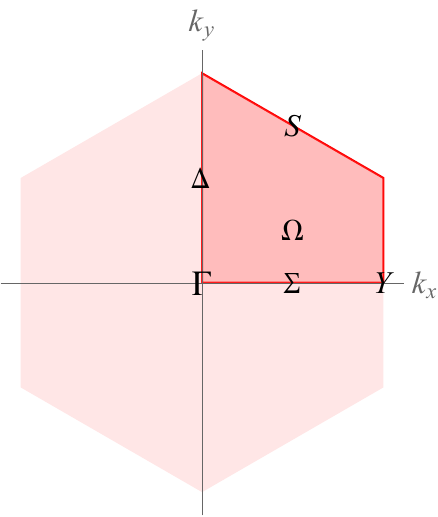}
    \caption{The Fundamental Domain, $\Omega$, for the first Brillouin zone of $C222$. The High Symmetry Points (HSPs) and Lines (HSLs) are labelled ${\Gamma, Y, S}$ and ${\Delta,\Sigma}$ respectively.}
    \label{fig:BZ}
\end{figure}

\noindent
 Whether a crystalline system's band structure is topologically non-trivial can be determined using the tools of Topological Quantum Chemistry (TQC) by simply knowing its space group, Wyckoff position, and orbital symmetry \cite{Bradlyn2017TopologicalChemistry, Cano2020BandChemistry}.
 We apply this method to a uniaxially strained Honeycomb lattice with non-degenerate $\{p_{x},p_{y}\}$ orbitals $\{C222, 4e ,(x,y)\}$. 
\noindent
The space group of the strained Honeycomb lattice is reduced from $P622$ to $C222$, and the point group at the nodes is reduced from $D_{3}$ to $C_{2}$.
\noindent
Firstly we introduce the Fundamental Domain of the hexagonal Brillouin zone with symmetry $C222$, as defined in Fig. \ref{fig:BZ}. The Fundamental Domain is the minimal region within the 1st Brillouin zone where points cannot be related via any symmetry operation belonging to the point group of the lattice \cite{Kruthoff2017TopologicalCombinatorics}. The band structure within the Fundamental Domain completely describes the band structure for the whole Brillouin zone.

\noindent
We can therefore completely describe the band structure using the High Symmetry Features (HSF) $\{\Gamma, Y, S,\Delta,\Sigma\}$ spanning the Fundamental domain as defined in Tab. \ref{tab:HSF}.  For each HSF, $\vec{k}$, there is a little co-group, $G_{\vec{k}}$, such that $\vec{k}$ is invariant under the set of symmetry operations belonging to this group up to a combination of reciprocal lattice translations $\vec{Q}=n_{1}\vec{b_{1}} + n_{2}\vec{b_{2}} + n_{3}\vec{b_{3}}$, if $n_{1},n_{2},n_{3} \in \mathbb{Z}$

\begin{table}[H]
\centering
\def\arraystretch{2}
\resizebox{0.4\textwidth}{!}{
\begin{tabular}{ |c|c|c|c|c| } 
 \hline
 {}& $\vec{k}$ & Little co-group, $G_{\vec{k}}$\\
 \hline
 $\Gamma$ & $(0,0)$ & $\{E,C_{2}(z),C_{2}(x),C_{2}(y)\}$ \\
 \hline
 Y & $(\frac{2\pi}{\sqrt{3}a},0)$ & $\{E,C_{2}(z),C_{2}(x),C_{2}(y)\}$\\ 
 \hline
 S & $(\frac{2\pi}{2\sqrt{3}a},\frac{\pi}{a})$ &  $\{E,C_{2}(z)\}$ \\ 
 \hline
 $\Delta$ & $(0,k_y)$ &  $\{E,C_{2}(y)\}$ \\ 
 \hline
 $\Sigma$ & $(k_x,0)$ &  $\{E,C_{2}(x)\}$ \\ 
 \hline
\end{tabular}}
\caption{HSF coordinates, $\vec{k}$, and little co-groups, $G_{\vec{k}}$, for space group $C222$. 
Note that the point groups for the HSFs $\{\Gamma, Y, S, \Delta, \Sigma\}$ are $\{D_{2}, D_{2}, C_{2}, C_{2}, C_{2}\}$ respectively. 
The associated rotation axes are indicated in parentheses after each operator. $a$ is the lattice constant.}
\label{tab:HSF}
\end{table}
\noindent
We must now consider how our lattice transforms under the little co-groups defined in Tab. \ref{tab:HSF} in the basis: $\{\psi_{A,p_{x}},\psi_{A,p_{y}},\psi_{B,p_{x}},\psi_{B,p_{y}}\}$. These representation for $g\in G_{\vec{k}}$ are written $\rho_{\vec{k}}(g)$. The three symmetry operations in this system, excluding the identity, act as a rotation by $\pi$ about the three Cartesian axes. The real Spherical Harmonics for the $p_{x}$ and $p_{y}$ orbitals are linear functions of $x$ and $y$ respectively.
\noindent
Additionally, we find $A$ and $B$ sites are switched under $C_{2}(z)$ and $C_{2}(y)$ rotations in real space.
\noindent
Additional momenta of some integer sum of reciprocal lattice vectors, $\Vec{Q}=(n_{1}\Vec{b}_1 + n_{2}\vec{b}_2)$, is required to conserve the momentum of the HSFs. In order to conserve momenta, upon rotation the state must pick up a phase factor $e^{-i\Vec{Q}\cdot\Vec{d}_i}$, where $\Vec{d}_i$ is the distance between sublattice $i$ and the axis of rotation, perpendicular to that axis. In the system $\{C222, 4e ,(x,y)\}$, we need only conserve the momentum on the $Y$ point under $C_2(z)$ 
 and $C_2(y)$ rotation by adding a phase factor of $e^{\pm i\frac{2 \pi}{3}}$ to sites A and B respectively.
 
\noindent
Given our system is periodic, 
we can represent the Hamiltonian as a sum of Bloch Hamiltonians, $H_{\vec{k}}$, in momentum space. $\rho_{\vec{k}}(g)$ commutes with the Bloch Hamiltonian: Therefore at a HSF the eigenstates of $H_{\vec{k}}$ must also be the eigenstates of $\rho_{\vec{k}}(g)$. From the perspective of band theory, this means the collection of bands below the Fermi level with momentum $\vec{k}$ corresponds to a representation of the little co-groups $G_{\vec{k}}$. This representation can be written as a sum of \emph{irreducible representations} (irreps), where one and two-dimensional irreps correspond to singular and degenerate bands respectively. In summary, in finding the irreps of each $G_{\vec{k}}$ we can enumerate the different combinations of irreps, and therefore determine the set of possible band structures \cite{Bercioux2017Topological2017, Cano2018BuildingRepresentations}. The representations are decomposed into $a_{j}$ irreps:

\begin{equation}
    a_{j} = \frac{1}{|G|}(\sum_{k} |C_{k}| \chi_i(C_k)\chi(C_k)^{*} )
\end{equation}
\noindent
Where $C_{k}$ represents the $k^{th}$ conjugacy class \footnote{Two elements, $a$ and $b$, are conjugate (and therefore belong to the same class) if there is an element $g$ in the group such that $b = gag^{-1}$. Note that the character of each element in a conjugacy class is the same.}. Using the character tables shown in Tab. \ref{tab:D2} and \ref{tab:C2}, we can decompose the representations in the little co-groups $\{G_{\Gamma},G_{Y}\}$ and $\{G_{S}, G_{\Delta},G_{\Sigma}\}$ respectively. The characters of the representations for each little co-group are found to be zero, excluding $E$. Therefore, decomposition is straightforward and all 1D irreps occur once for little co-groups isomorphic to $D_{2}$ and twice for those isomorphic to $C_2$, as shown in Tab. \ref{tab:irreps}.

\begin{table}[H]
$$
\begin{array}{c|rrrr}
 &E&C_{2}(z)&C_{2}(y)&C_{2}(x)	\cr
\hline
\Gamma_{1} (A_{1})	&1&	1&	1	&1	\cr
\Gamma_{2} (B_{1}) &1&	1&	-1&	-1	\cr
\Gamma_{3}	(B_{3}) &1&	-1&	-1&	1	\cr
\Gamma_{4}	(B_{2}) &1&	-1&	1&	-1	\cr
\end{array}
$$
\caption{Character table for little co-groups $\{G_{\Gamma},G_{Y}\}\approx D_2$}
\label{tab:D2}
\end{table}

\begin{table}[H]
$$
\begin{array}{c|rr}
&E&C_{2}	\cr
\hline
\Gamma_{1} (A)	&1&	1	\cr
\Gamma_{2} (B)	&1&	-1	\cr
\end{array}
$$
\caption{Character table for little co-groups $\{G_{S},G_{\Delta},G_{\Sigma}\}\approx C_2$}
\label{tab:C2}
\end{table}

\begin{table}[H]
\centering
\def\arraystretch{2}
\resizebox{0.9\textwidth}{!}{
\begin{tabular}{|c|c|c|c|c|c|}
 \hline
 BR & $\Gamma$ & $Y$ & $S$ & $\Delta$ & $\Sigma$ \\
 \hline
 $\rho \uparrow G$ & 
 $\Gamma_1 \oplus \Gamma_2 \oplus \Gamma_3 \oplus \Gamma_4$ & 
 $Y_{1} \oplus Y_{2} \oplus Y_{3} \oplus Y_{4}$ &
 $2S_{1} \oplus 2S_{2}$ &
 $2\Delta_{1} \oplus 2\Delta_{2}$ &
 $2\Sigma_{1} \oplus 2\Sigma_{2}$ \\
 \hline
\end{tabular}
}
\caption{Irreducible representations of the little co-groups 
$\{G_{\Gamma}, G_{Y}, G_{S}, G_{\Delta}, G_{\Sigma}\}$ 
induced from $\{p_{x}, p_{y}\}$ orbitals on a uniaxially strained Honeycomb lattice. 
The irrep indices are defined in character tables 
Tab.~\ref{tab:D2} and Tab.~\ref{tab:C2}.}
\label{tab:irreps}
\end{table}

\subsection{Compatibility Relations}
\noindent
Imagining there are no symmetry constraints, given that there are 4 distinct irreps at the $\Gamma$ and $Y$ points and 2 distinct irreps at the $S$ point, there would be $3456$ distinct combinations of the irreps defined in Tab. \ref{tab:irreps}. However, there are symmetry constraints, and these come in the form of compatibility relations. Using Tab. \ref{tab:HSF}, \ref{tab:D2}, \ref{tab:C2}, \ref{tab:irreps}, we construct the compatibility relations connecting $\Gamma$ and $Y$ along lines $\Delta$ and $\Sigma$, which reduce to a pair of independent linear equations:

\begin{subequations}
    \begin{equation}
        n_{3}^{\Gamma} -  n_{4}^{\Gamma} = n_{3}^{Y} -  n_{4}^{Y}
    \end{equation}
    \begin{equation}
        n_{2}^{\Gamma} -  n_{1}^{\Gamma} = n_{2}^{Y} -  n_{1}^{Y}
    \end{equation}
\label{eq:compat}
\end{subequations}

\noindent
Where $n_{i}^{\vec{k}}$ is the number of bands at $\vec{k}$ transforming under irrep $\Gamma_{i}$. Additionally there is the trivial restriction of $n_{1}^{\Gamma}+n_{2}^{\Gamma}+n_{3}^{\Gamma}+n_{4}^{\Gamma}=n_{1}^{Y}+n_{2}^{Y}+n_{3}^{Y}+n_{4}^{Y}=N$
, where $N$ is the total number of valance bands. Note that we cannot place any restrictions on HSP $S$ as it is a \emph{floating} point, i.e. there are no HSLs connecting it to other HSPs, and so is not included in the compatibility relations. Therefore the number of distinct band structures is reduced from 3456 to 14.

\subsection{Topological insulating bands}
For a given lattice symmetry, at the atomic limit\footnote{Where the sites are separated until there is no hopping between sites.} the bands are in one of a finite number of possible Elementary Band Representations (EBRs), as enumerated in the Bilbao Crystallographic Server \cite{Aroyo2006BilbaoPrograms,Aroyo2006BilbaoGroups}. A topological band is defined as one which cannot be continuously deformed to an atomic limit insulating state without closing a gap. We therefore say that a set of bands which can't be represented as a positive sum of EBRs are topologically non-trivial. These EBRs are defined in the basis of data symmetry vector;

\begin{equation}
    \vec{v}=(n_{1}^{\Gamma},n_{2}^{\Gamma},n_{3}^{\Gamma},n_{4}^{\Gamma},n_{1}^{Y},n_{2}^{Y},n_{3}^{Y},n_{4}^{Y},n_{1}^{S},n_{2}^{S},N ).
\end{equation}

Therefore EBRs can be summed using simple linear algebra.
\noindent
The EBRs at the Maximal Wyckoff Position (MWP), $2a:(0,0)$, $2b:(1/2,0)$ and $4k:(1/4,1/4)$ are defined in Eq. \ref{eq:2a}, \ref{eq:2b} and \ref{eq:4k} respectively:

\vspace{20pt}

\begin{minipage}{0.5\textwidth}
    \begin{subequations}
        \begin{equation}
             \vec{e}_{2a,A_{1}} = (1,0,0,0,1,0,0,0,1,0,1)
        \end{equation}
                \begin{equation}
            \vec{e}_{2a,B_{1}} = (0,1,0,0,0,1,0,0,1,0,1)
        \end{equation}
                \begin{equation}
            \vec{e}_{2a,B_{3}} = (0,0,1,0,0,0,1,0,0,1,1) 
        \end{equation}
                \begin{equation}
             \vec{e}_{2a,B_{2}} = (0,0,0,1,0,0,0,1,0,1,1) 
        \end{equation}
    \label{eq:2a}
    \end{subequations}
    \end{minipage}\hfill
\begin{minipage}{0.5\textwidth}
    \begin{subequations}
            \begin{equation}
             \vec{e}_{2b,A_{1}} = (1,0,0,0,1,0,0,0,0,1,1)
        \end{equation}
                \begin{equation}
             \vec{e}_{2b,B_{1}} = (0,1,0,0,0,1,0,0,0,1,1)
        \end{equation}
                \begin{equation}
           \vec{e}_{2b,B_{3}} = (0,0,1,0,0,0,1,0,1,0,1)
        \end{equation}
                \begin{equation}
             \vec{e}_{2b,B_{2}} = (0,0,0,1,0,0,0,1,1,0,1)
        \end{equation}
     \label{eq:2b}
    \end{subequations}
\end{minipage}

\begin{subequations}
    \begin{gather}
         \vec{e}_{4k,A} = (1,1,0,0,0,0,1,1,1,1,2)\\
         \vec{e}_{4k,B} = (0,0,1,1,1,1,0,0,1,1,2) 
        \end{gather}
        \label{eq:4k}
\end{subequations}

Note that the EBRs of MWP $2b:(1/2,0)$ differs from $2a:(0,0)$ by values of $n_{1}^{S}$ and $n_{2}^{S}$.

 \begin{figure}[H]
    \centering
    \includegraphics[width=0.6\textwidth]{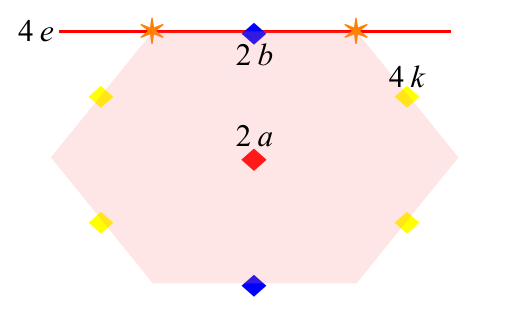}
    \caption{MWPs, $2a:(0,0)$, $2b:(1/2,0)$ and $4k:(1/4,1/4)$, of space group $C222$. Orbitals are located along non-MWP $4e:(x,0)$.}
    \label{fig:WP}
\end{figure}

\noindent
Given the compatibility relations defined in Eq. \ref{eq:compat}, when considering isolated bands, i.e. $N=1$, it is only possible to obtain EBR Eq. \ref{eq:2a} and \ref{eq:2b}, therefore telling us there are no isolated topological bands. 
\noindent
This also tells us that given $N=4$ is a topologically trivial set of bands, $N=3$ is also a topologically trivial set. 

\noindent
For $N=2$ any positive combination of Eq. \ref{eq:2a} and \ref{eq:2b} are possible, and Eq. \ref{eq:4k} alone. Remaining are 20 pairs of non-trivial bands, 14 of which are distinct. However only four of 14 pairs of bands are possible given restrictions from the Hamiltonian discussed in Appendix B. 

    \begin{subequations}
        \begin{equation}
         \vec{v}_{TI,\MakeUppercase{\romannumeral 1}} = (1,1,0,0,0,0,1,1,0,2,2)
        \end{equation}
        \begin{equation}
         \vec{v}_{TI,\MakeUppercase{\romannumeral 2}} = (1,1,0,0,0,0,1,1,2,0,2)
        \end{equation}
    \begin{equation}
         \vec{v}_{TI,\MakeUppercase{\romannumeral 3}} = (0,0,1,1,1,1,0,0,0,2,2)
    \end{equation}
    \begin{equation}
         \vec{v}_{TI,\MakeUppercase{\romannumeral 4}} = (0,0,1,1,1,1,0,0,2,0,2)
    \end{equation}
    \label{eq:TI}
    \end{subequations}

\vspace{1cm}

\subsubsection{Obstructed Atomic Insulating bands}
\noindent
A system is in an obstructed atomic insulating phase when the orbital positions are different to the MWPs, such that any deformation to the atomic limit which respects the symmetry of the system is inhibited \cite{Benalcazar2017QuantizedInsulators,Schindler2018Higher-orderInsulators}. Examining Fig. \ref{fig:WP} we see that it is possible for two orbitals (along Wyckoff position $4e:(x,0)$) to converge without obstruction to a sum of $2b:(1/2,0)$ EBRs. This is however not possible for $2a:(0,0)$ and $4k:(1/4,1/4)$, meaning any band which is formed from a sum of EBRs at these positions is obstructed. Therefore Eq. \ref{eq:2a} and Eq. \ref{eq:4k} are obstructed atomic bands, and any combination of Eq. \ref{eq:2a} with itself are also obstructed atomic bands, making 16 in total. We are only interested in those defined in Eq. \ref{eq:4k}, again given restrictions from the Hamiltonian discussed in Appendix B.

\begin{subequations}
    \begin{equation}
         \vec{v}_{HOTI,\MakeUppercase{\romannumeral 1}} = (1,1,0,0,0,0,1,1,1,1,2)
    \end{equation}
    \begin{equation}
         \vec{v}_{HOTI,\MakeUppercase{\romannumeral 2}} = (0,0,1,1,1,1,0,0,1,1,2)
    \end{equation}
    \label{eq:HOTI}
\end{subequations}

\noindent
HOTI phases fall into the category of obstructed atomic insulating phases, and can be classified by the presence of fractionalized corner states. We find at half filling charge is distributed at $\pm e/2$ either end of a strained Honeycomb lattice on a $C222$ slab.

\section{Appendix B: Topological Domain Boundaries}

To determine whether the topological bands in Eq. \ref{eq:TI} and \ref{eq:HOTI} exist, the energetic ordering of the irreps at HSPs is required, and therefore we must consider the H-XY Hamiltonian. H-XY is transformed into the basis of irreps at each HSP $\{\Gamma_{1},\Gamma_{2},\Gamma_{3},\Gamma_{4}\}$, $\{Y_{1},Y_{2},Y_{3},Y_{4}\}$, $\{S_{1(1)},S_{1(2)},S_{2(1)},S_{2(2)}\}$. This is done using the projection operator for each HSP:

\begin{equation}
    P(\Gamma) = \sum_i \chi(\Gamma(G_{i}))G_{i}
\end{equation}
\noindent
For points $\Gamma$ and $Y$, given there are four distinct one-dimensional irreps, the Hamiltonian is fully diagonalized. While for point $S$, as there are copies of the same irrep, these states will mix, meaning the Hamiltonian is diagonalized into two blocks. The eigenvalues of these irreps are:

\begin{minipage}{0.5\textwidth}
\begin{subequations}
    \begin{equation}
        E_{1}^{\Gamma}=E_{x}-\frac{1}{2}( 3t_{\pi_{0}}+ t_{\sigma_{0}})-t_{\sigma}
    \end{equation}
    \begin{equation}
        E_{2}^{\Gamma}=E_{y}-\frac{1}{2}(t_{\pi_{0}}+3 t_{\sigma_{0}})-t_{\pi}
    \end{equation}
    \begin{equation}
        E_{3}^{\Gamma}=E_{x}+\frac{1}{2}( 3t_{\pi_{0}}+ t_{\sigma_{0}})+t_{\sigma}
    \end{equation}
    \begin{equation}
        E_{4}^{\Gamma}=E_{y}+\frac{1}{2}( t_{\pi_{0}}+3 t_{\sigma_{0}})+t_{\pi}
    \end{equation}
\end{subequations}
\end{minipage}
\begin{minipage}{0.5\textwidth}
    \begin{subequations}
    \begin{equation}
        E_{1}^{Y}=E_{x}+\frac{1}{2}( 3t_{\pi_{0}}+ t_{\sigma_{0}})-t_\sigma
    \end{equation}
    \begin{equation}
        E_{2}^{Y}=E_{y}+\frac{1}{2}( t_{\pi_{0}}+3 t_{\sigma_{0}})-t_\pi
    \end{equation}
    \begin{equation}
        E_{3}^{Y}=E_{x}-\frac{1}{2}( 3t_{\pi_{0}}+ t_{\sigma_{0}})+t_\sigma
    \end{equation}
    \begin{equation}
        E_{4}^{Y}=E_{y}-\frac{1}{2}( t_{\pi_{0}}+3 t_{\sigma_{0}})+t_\pi
    \end{equation}
    \end{subequations}
\end{minipage}
\begin{minipage}{1\textwidth}
\begin{subequations}
\begin{equation}
        E_{1(1)}^{S}=\frac{1}{2} \left(E_{x}+E_{y}-t_{\pi}-t_{\sigma}+\sqrt{(E_{x}-E_{y}+t_{\pi}-t_{\sigma})^2+3 (t_{\pi_{0}}-t_{\sigma_{0}})^2} \right)
    \end{equation}
    \begin{equation}
        E_{1(2)}^{S}=\frac{1}{2} \left(E_{x}+E_{y}-t_{\pi}-t_{\sigma} \\ -\sqrt{(E_{x}-E_{y}+t_{\pi}-t_{\sigma})^2+3 (t_{\pi_{0}}-t_{\sigma_{0}})^2} \right)
    \end{equation}
    \begin{equation}
        E_{2(1)}^{S}=\frac{1}{2} \left(E_{x}+E_{y}+t_{\pi}+t_{\sigma}+\sqrt{(E_{x}-E_{y}-t_{\pi}+t_{\sigma})^2+3 (t_{\pi_{0}}-t_{\sigma_{0}})^2} \right)
    \end{equation}
    \begin{equation}
        E_{2(2)}^{S}=\frac{1}{2} \left(E_{x}+E_{y}+t_{\pi}+t_{\sigma}-\sqrt{(E_{x}-E_{y}-t_{\pi}+t_{\sigma})^2+3 (t_{\pi_{0}}-t_{\sigma_{0}})^2} \right)
\end{equation}
\end{subequations}
\end{minipage}

\noindent
Where $E_{i}^{\vec k}$ is the energy of a band transforming under irrep $\Gamma_{i}$ at high symmetry point $\vec k$.

\section{Appendix C: Computational Detail}

The frontier molecular orbitals (MOs) of the monomer units were calculated in Gaussian 16 and visualised using GaussView software \cite{M.J.FrischandG.W.TrucksandH.B.SchlegelandG.E.Scuseria2016GaussianC.01, RoyDenningtonandToddA.KeithandJohnM.Millam2019GaussView6}. The monomers geometries were optimised in the ground state, before the MOs were found at the B3LYP 6-31G level.

\noindent
The COF structures were relaxed in two steps in the Vienna ab initio simulation package (VASP) \cite{Kresse1996EfficientSet}: Firstly just the atomic positions were relaxed with a fixed unit cell (ISIF = 2) and no symmetry restriction (ISYM = 0), and then the unit cell was relaxed (ISIF = 3) with symmetry restrictions (ISYM = 2). We used a $\Gamma$ centred Monkhort-Pack k-point mesh of size $2\times2\times1$ for CTF-2 and $1\times1\times1$ for the uniaxially strained CTF-2. All the atoms were allowed to relax until the atomic forces were smaller than 0.02 $\ \si{\electronvolt}/$\AA. All the geometries of the 2D COFs were optimised in the ground state.

\noindent
For the Self-Consistent Field (SCF) step the density of k-point mesh was increased to $3\times3\times1$ for CTF-2, and remains the same for uniaxially strained CTF-2. A vacuum layer over $15$\AA~ thick was used to ensure electronic decoupling between neighbouring slabs. Dispersion corrections of the van der Waals interactions (DFT-D2) were considered in the calculations on multi-layer thin films. \\

\noindent
The plane-wave cutoff energy was always set to 500 \si{\electronvolt}.
The band structure of all COFs were calculated using a Perdew-Burke-Ernzerhof (PBE) Generalized Gradient Approximation (GGA) functional \cite{Perdew1996GeneralizedSimple, Wang2021VASPKIT:Code}. All other parameters were set to their default VASP values unless stated otherwise.

\begin{figure}[H]
\centering
\begin{overpic}[width=1\textwidth]{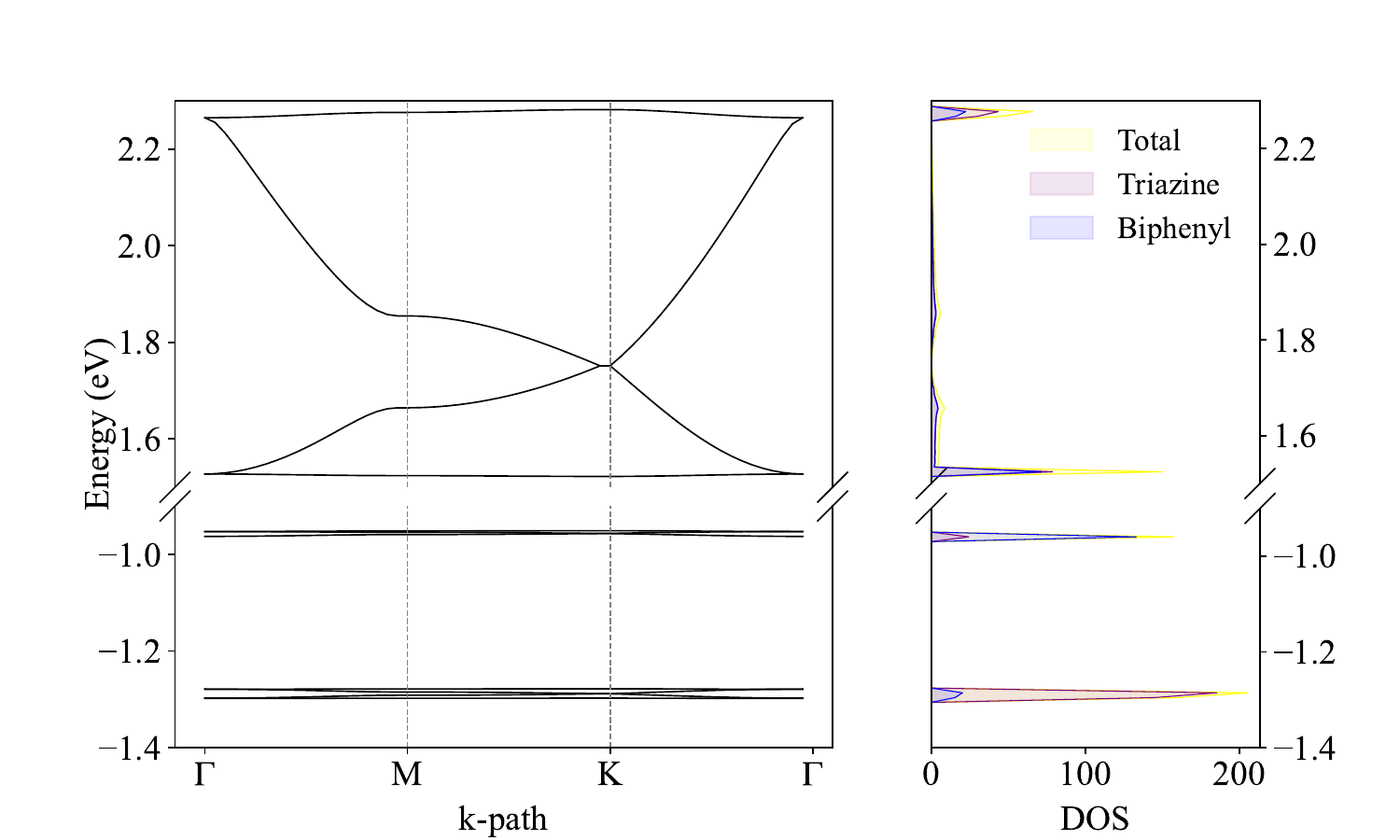}
  \put(67,28){\includegraphics[width=0.22\textwidth, trim={2.5cm 9cm 3.2cm 9cm}, clip]{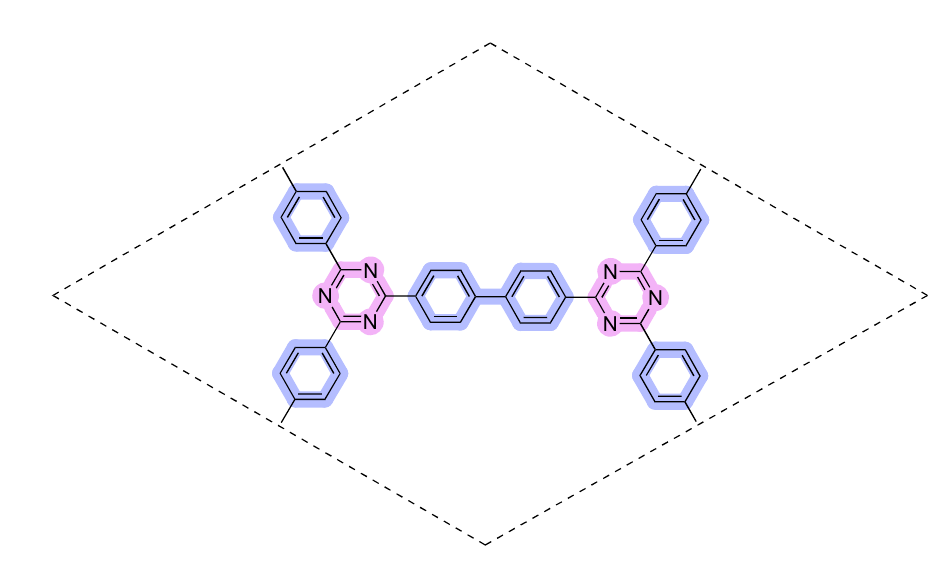}}
  \put(13,9){%
    \begin{tikzpicture}[scale=0.5]
      \draw[white, line width=0pt, fill=yellow, fill opacity=0.2] (0,0) rectangle (15,1);
    \end{tikzpicture}
  }
\end{overpic}
            \caption{{\it Ab initio} band structure and projected partial density of states (PDOS) of core and linker units of CTF-2. PDOS are plotted along the energy axis to indicate contribution of different parts of the unit cell to the bands. The relevant bands for the H-XY model are located just below -1.2 eV. Lattice parameters $a=b=22.06\text{\AA}$ and space group $P622$. The Fermi level is  -4.2099$\ \si{\electronvolt}$ and the Bandgap is 2.4734$\ \si{\electronvolt}$.}
\end{figure}

\begin{figure}[H]
\centering
\begin{overpic}[width=1\textwidth]{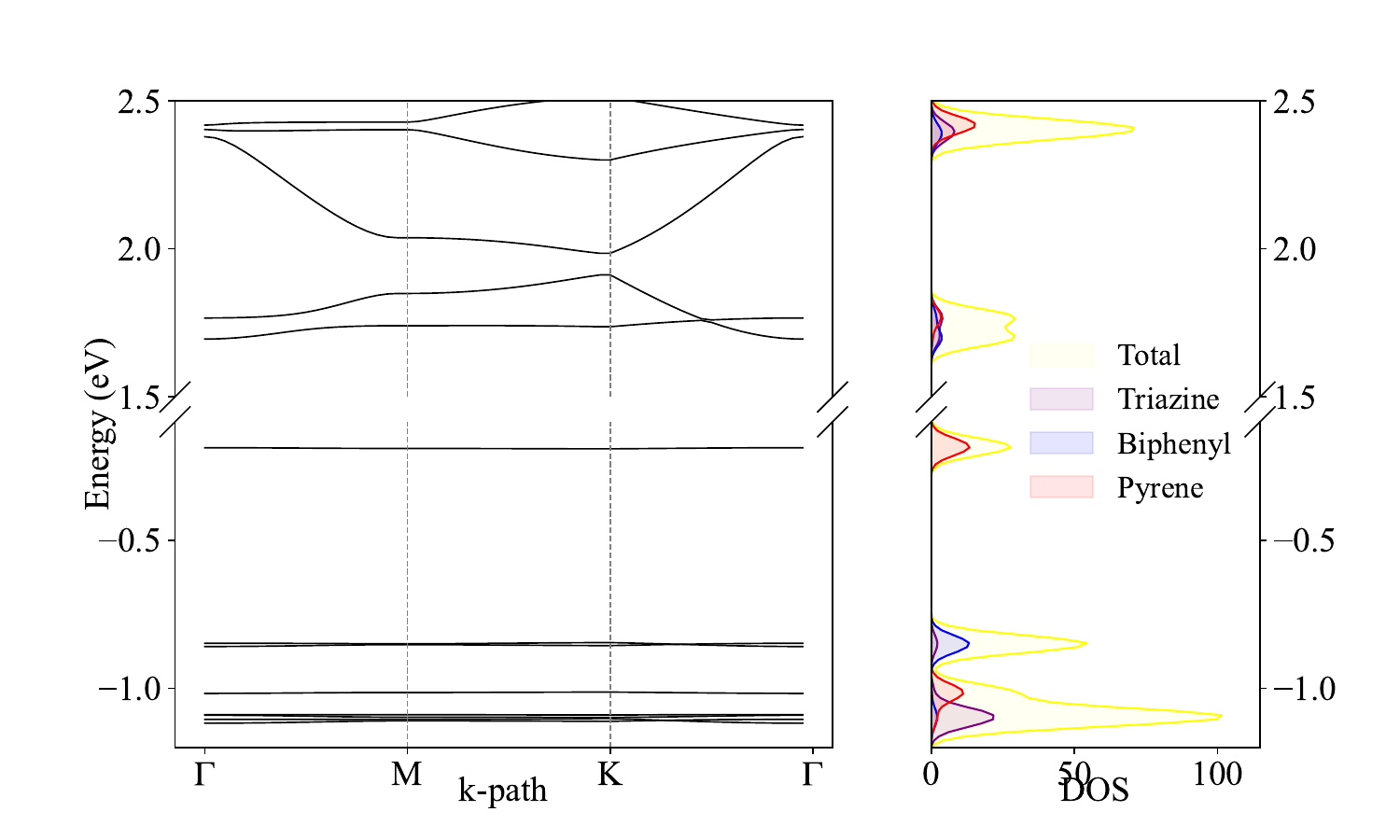}
  \put(71,39){\includegraphics[width=0.18\textwidth, trim={3.4cm 9.7cm 3.4cm 9.7cm}, clip]{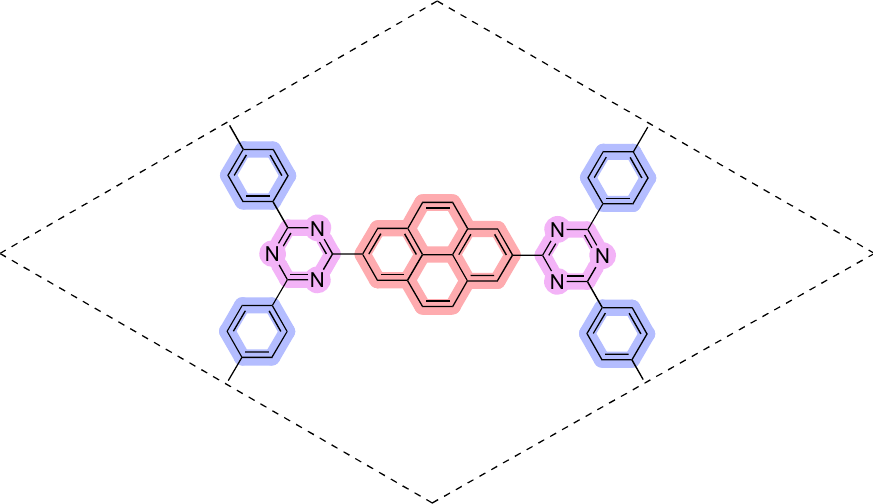}}
  \put(14,7){%
    \begin{tikzpicture}[scale=0.5]
      \draw[white,line width=0pt, fill=yellow, fill opacity=0.2] (0,0) rectangle (14.5,1);
    \end{tikzpicture}
  }

\end{overpic}
    \caption{{\it Ab initio} band structure and projected PDOS of core and linker units of uniaxially strained CTF-2, obtained by replacing biphenyl linkers in one-direction with pyrene. PDOS are plotted along the energy axis to indicate contribution of different parts of the unit cell to the bands. 
    The relevant bands for the uniaxially strained H-XY model are located just below -1.0 eV. Lattice parameters $a=b=22.43$\text{\AA} and space group $C222$}
    \label{fig:strained_bands_pdos}
\end{figure}

\section{Appendix D: Other Relevant Space Groups}

Note that there are a number of space groups where action on $p_x$, $p_y$ orbitals is the same as $C222$, so the H-XY model remains appropriate and the TQC analysis of the bands remains unchanged. These are enumerate in Tab. \ref{tab:spacegroups}.

\begin{table}[H]
\centering
\def\arraystretch{1}
\resizebox{0.6\textwidth}{!}{
\begin{tabular}{|c|c|c|c|}
 \hline
 SG & WP x-strain & WP y-strain & OS \\
 \hline
 C222$_1$ & 4a & 4b & (x,y) \\
 C222 & 4e & 4g & (x,y) \\
 Cmm2 & 4d & 4e & (x,y) \\
 Cmc2$_1$ &  & 4a & (x,y) \\
 Cmcm &  & 4b & (x,y) \\
 Cmmm & 4g & 4i & (x,y) \\
 \hline
\end{tabular}
}
\caption{Space groups with equivalent symmetry analysis to $C222$ for $p_x$, $p_y$ orbitals.}
\label{tab:spacegroups}
\end{table}

\end{document}